\documentclass[aps,prd,reprint,floatfix,superscriptaddress,showkeys,nofootinbib]{revtex4-1}

\setcitestyle{authoryear,round}
\usepackage[utf8]{inputenc}
\usepackage[T1]{fontenc}

\usepackage{graphicx}

\usepackage{siunitx}
\usepackage{mathtools}
\usepackage{amsfonts}
\usepackage{bm}
\usepackage{bbm}

\usepackage{physics}

\usepackage{verbatim}
\usepackage{rotate}
\usepackage{color}
\usepackage{aas_macros}
\usepackage{ulem}
\DeclareFontFamily{OT1}{pzc}{}
\DeclareFontShape{OT1}{pzc}{m}{it}%
            {<-> s * [1.10] pzcmi7t}{}
\DeclareMathAlphabet{\mathscr}{OT1}{pzc}%
                                {m}{it}

\definecolor{RedWine}{rgb}{0.743,0,0}
\definecolor{DarkPastelGreen}{rgb}{0.01,0.75,0.24}
\definecolor{green(pigment)}{rgb}{0.0,0.65,0.31}
\definecolor{RoyalBlue}{rgb}{0.25,0.41,0.88}

\newcommand{\be}{\begin{equation}}
\newcommand{\ee}{\end{equation}}
\newcommand{\bea}{\begin{eqnarray}}
\newcommand{\eea}{\end{eqnarray}}
\def\ba#1\ea{\begin{align}#1\end{align}}

\def\({\left(}
\def\){\right)}
\def\<{\left\langle}
\def\>{\right\rangle}

\newcommand{\vs}{\nonumber\\}

\def\vr{{\bm{r}}}
\def\vx{{\bm{x}}}

\def\xhat{{\hat{\bm{x}}}}

\def\rhat{{\hat{\bm{r}}}}

\def\nbar{{\bar{n}}}

\def\fnl{f_\mathrm{NL}}
\def\fnlloc{f_{\mathrm{NL}}^{loc}}

\DeclareSIUnit \parsec {pc}
\DeclareSIUnit \h {\text{$h$}}
\DeclareSIUnit \year {yr}
\DeclareSIUnit \solarmass {M_\odot}
\DeclareSIUnit \Mpc {\mega\parsec}

\def\min{\mathrm{min}}
\def\max{\mathrm{max}}

\def\bx{\bold{x}}

\def\dd{\mathrm{d}}

\newcommand{\ignore}[1]{}

\def\myapp#1#2{%
  \mathrel{%
    \setbox0=\hbox{$#1\sim$}%
    \setbox2=\hbox{%
      \rlap{\hbox{$#1\propto$}}%
      \lower1.1\ht0\box0%
    }%
    \raise0.25\ht2\box2%
  }%
}

\newcommand{\incgraph}[2][0.49]{\includegraphics[width=#1\textwidth]{#2}}

\usepackage[colorlinks]{hyperref}
\usepackage[capitalise]{cleveref}

\hypersetup{
colorlinks=true,
citecolor=cyan,}

\begin{document}

\title{Demonstrating the Use of the Spherical Fourier Bessel Basis for Large Scale Clustering Systematics Discovery and Mitigation with eBOSS}


\author{Sean \surname{Bruton}}
\email{sbruton@caltech.edu}
\affiliation{California Institute of Technology, Pasadena, CA 91125, USA}

\author{James R. \surname{Cheshire} IV}
\affiliation{California Institute of Technology, Pasadena, CA 91125, USA}

\author{Olivier \surname{Dor\'{e}}}
\affiliation{California Institute of Technology, Pasadena, CA 91125, USA}
\affiliation{Jet Propulsion Laboratory, California Institute of Technology, Pasadena, CA 91109, USA}

\author{Henry S. \surname{Grasshorn Gebhardt}}
\affiliation{California Institute of Technology, Pasadena, CA 91125, USA}
\affiliation{Jet Propulsion Laboratory, California Institute of Technology, Pasadena, CA 91109, USA}

\author{Robin Y. \surname{Wen}}
\affiliation{California Institute of Technology, Pasadena, CA 91125, USA}

\begin{abstract}
The spherical Fourier-Bessel (SFB) basis, in separating the angular and radial modes of the power spectrum, permits a targeted identification and mitigation of systematics in clustering surveys while retaining more cosmological signal than traditional bases. We demonstrate this principle on the eBOSS DR16 luminous red galaxy (LRG) and quasistellar object (QSO) samples, identifying modes which may be contaminated by systematics. Our initial inference on the LRG  sample yields an $\fnl$ value consistent with zero, while the QSO value is in slight tension with zero. Using the SFB basis, we vary the selection of angular and radial modes to search for inconsistencies in the inferred value of $\fnl$, an indicator of underlying systematics. In the QSO sample, we find evidence ($p \lesssim 0.002$ compared to the same cuts on EZMocks) of a systematic afflicting large physical scales, which is consistent with residual stellar contamination; we also find evidence for an unknown systematic in the QSO sample ($p=0.01$), with consistent but less significant evidence in the LRGs ($p\approx0.06$), at the approximate angular plate and imaging scale of eBOSS. 
\end{abstract}


\maketitle

\section{Introduction}
\label{sec:intro}
Ongoing and imminent galaxy surveys, such as DESI \citep{Adame+:2025JCAP...07..017A}, SPHEREx \citep{Dore+:2014arXiv1412.4872D,Bock+:2026ApJ...999..139B},
Rubin/LSST \citep{Ivezic+:2019ApJ...873..111I}, Euclid \citep{EuclidCollaboration:2025A&A...697A...1E},
and Roman \citep{Eifler+:2021MNRAS.507.1746E} are or will soon be measuring the 3-dimensional large-scale structure
(LSS) of the Universe with unprecedented statistical precision. Advances in
technology, infrastructure, and in observational and computational techniques
mean that these surveys will measure $\mathcal{O}(10^9)$ galaxies over an
unparalleled cosmological volume. With such advances in statistical precision,
however, small subtleties in modeling or contamination from various sources of systematic error will dominate uncertainties and demand detailed understanding.

One key goal of the latest generation of galaxy surveys is to measure the
deviation of the spectrum of primordial perturbations from that of a normal
distribution; so-called ``primordial non-Gaussianity''. Local-type
non-Gaussianity, parameterized by $\fnlloc$ (hereafter $\fnl$) \citep{Komatsu+:2002nmgm.meet.2009K, Dalal+:2008PhRvD..77l3514D}, will be particularly
well-constrained with these datasets. The current leading constraint comes from the
cosmic microwave background (CMB) bispectrum results from the
\textit{Planck} Collaboration \citep{PlanckCollaboration:2020A&A...641A...9P}, $\fnl = 0.9 \pm 5.1$. However,
inherently 2-dimensional CMB observations are nearly limited by cosmic
variance, suggesting 3-D surveys of large-scale structure as the natural avenues
to furthering constraints on $\fnl$. A challenge is that accurate constraints on
$\fnl$ from large-scale structure require uncontaminated measurements of galaxy clustering on very large scales. These large scales are
those which are particularly severely affected by various observational and
instrumental systematic effects, necessitating careful characterization and
mitigation.

Various approaches have been used to mitigate systematic effects in LSS surveys,
with some analyses leveraging a combination of strategies.
Full simulation of systematic effects and marginalizing over them in the
inference step (e.g. \citet{Modi+:2019JCAP...11..023M}) can be a powerful technique for systematic effects which are
themselves well modeled, but can be computationally expensive. Another common
technique utilizes systematic template deprojection, where template maps are regressed against the observed galaxy density fields to
correct for the modulation of the density field by the contaminant \citep{Weaverdyck+:2021MNRAS.503.5061W}.
Both of these approaches rely on understanding the landscape of and the modeling
of individual systematic effects, thus different approaches must be used to
identify and mitigate unknown or unmodeled systematic effects. Unknown systematics can be probed by injecting synthetic sources with known properties and propagating
them through the analysis procedure, e.g. \citet{Everett+:2022ApJS..258...15E}. This can yield a
measurement of the transfer function accounting for many systematics, though it is also computationally
expensive, and relies on the accurate simulation 
of any instrumental effects. Yet
another common technique, which will be the focus in this paper, is the explicit
removal (or down-weighting) of specific scales/modes which are suspected of
being systematically contaminated. These may often be identified through various
data splits and consistency tests.

This latter ``mode-cutting'' approach is typically a powerful way of ensuring
that the resulting dataset is uncontaminated, but may also remove significant
amounts of cosmological signal. In \citet{Chaussidon+:2025JCAP...06..029C}, extensive
template-based systematic regression is applied, however significant differences
in the measured power spectrum are still found between different observing
regions for the LRG sample after reweighting to account for the effects
captured by the templates, suggesting remaining un-modeled systematic
contamination at large scales. This was addressed by increasing the minimum
wave number $k_{\rm min}$ from $0.003~h$Mpc$^{-1}$ to $0.006~h$Mpc$^{-1}$, though at
the cost of a $\sim30\%$ degradation in constraints on $\fnl$ for the LRG sample alone. Unlike the cosmological signal, systematics, e.g. stars \citep{Wen+:2025PhRvD.112}, are generally not homogeneous and isotropic. Due to their association with the observer position, systematics tend to be associated with a spherical geometry; however, typical power spectrum estimators are not naturally suited for the separation of angular and radial effects. Because of this, power spectrum estimators (e.g. \citet{Yamamoto+:2006PASJ...58...93Y}) entangle angular and radial information and any associated cuts (e.g., cuts on $k_{\rm min}$) remove both systematics and signal, when a more natural geometry allows one to cut systematics in an isolated way.

The spherical Fourier-Bessel (SFB) formalism described in \S\ref{sec:sfb_basis}
overcomes many of these deficiencies by matching the geometry of the estimator
to the intrinsic geometry of realistic galaxy surveys: that of an observer
situated at the center of a spherical geometry. By working in a spherical basis,
wide-angle effects that arise from flat-sky approximations in Cartesian estimators are naturally captured, and effects related to observation
rather than to cosmology (i.e., systematics) are more easily distinguished. The
inherent 3-dimensional nature of the SFB basis, its large number of available modes, and
its alignment with the geometry of observation make it a powerful framework with
which to conduct cosmological analyses of large-scale structure, and
particularly suit it to the mode-cutting approach to systematics mitigation.
Systematics will be both more readily identified and less entangled with
cosmology in this basis, meaning more precise cuts can be made with minimized
loss of useful cosmological information. The effectiveness of mode cutting at mitigating
systematics in the context of the SFB formalism was recently demonstrated in \citet{Wen+:2025PhRvD.112}.

The \texttt{SuperFaB}\footnote{https://github.com/hsgg/SphericalFourierBesselDecompositions.jl} SFB estimator was presented in \citet{GrasshornGebhardt+:2021PhRvD.104l3548G} and validated on a
variety of simulations (including both ``complete'' and ``realistic'' EZmock
simulations \citep{Zhao+:2021MNRAS.503.1149Z}) in
\citet{GrasshornGebhardt+:2024PhRvD.109h3502G}. Using a covariance constructed
from 1000 EZmock realizations, the \texttt{SuperFaB} estimator was able to obtain
unbiased constraints on $\fnl$ as measured from 100 realizations.

This paper builds upon the EZmock validation result by applying the
\texttt{SuperFaB} estimator to the extended Baryon Oscillation Spectroscopic
Survey (eBOSS) DR16 luminous red galaxy (LRG) and
quasistellar object (QSO) samples \citep{Ross+:2020MNRAS.498.2354R} to produce
constraints on $\fnl$, before proceeding to apply the
results of \citet{Wen+:2025PhRvD.112} in attempting to cut modes which are
suspected to be contaminated. Previous efforts to constrain $\fnl$ using
standard Fourier-space power spectrum multipole analyses applied to
eBOSS QSOs include \citet{Castorina+:2019JCAP...09..010C} (using the DR14 release),
\citet{Qin+:2019MNRAS.487.5235Q}, and \citet{Cagliari+:2024JCAP...08..036C} (using the DR16
release). \citet{Cagliari+:2024JCAP...08..036C} obtain a constraint
$-4 < \fnl < 27$ at $68\%$ confidence, though they themselves note a
suspicion of residual systematic contamination based on the suboptimality of the
constraint as compared to Fisher matrix-based forecasts.
\citet{Cagliari+:2025JCAP...07..043C} obtain a constraint $-6 < \fnl < 20$ at
$68\%$ confidence by using the Fourier-space bispectrum as well as the power
spectrum to analyze the eBOSS DR16 QSO sample, noting that these results are
consistent with expectations. The eBOSS LRG sample has been less used for $\fnl$ analyses\footnote{The eBOSS LRG sample covers a smaller effective volume than the BOSS LRG sample due to its more limited sky area.}, while the BOSS DR12 galaxy sample, which mostly consists of LRGs, has been analyzed by both \citet{D'Amico+:2025PhRvD.111f3514D} and \citet{Cabass+:2022PhRvD.106d3506C} using the bispectrum in combination with the power spectrum.

For clarity, throughout this paper we define a ``purely angular'' systematic as one which
solely \emph{varies} angularly, i.e., where the modulation of the density field
is purely a function of angle, $\Delta\delta(\vx) \sim f(\xhat)$, where
$f(\xhat)$ is some function of the normal vector $\xhat$. 
Conversely, we define a ``purely radial'' systematic as one which
solely \emph{varies} radially, i.e., $\Delta\delta(\vx) \sim C f(x)$, where $x$
denotes the magnitude of $\bx$.

This paper is organized as follows: In \S\ref{sec:sfb_basis}, we review the
spherical Fourier-Bessel formalism, and the modeling of relevant observational effects
and of the galaxy-halo bias. In
\S\ref{sec:fnl_inference}, we describe the eBOSS LRG and
QSO datasets and present initial SFB-derived parameter constraints
with the fiducial mode selection. In \S\ref{sec:cleaning}, we describe the
motivation, procedure and results of the systematic-cleaning mode-cutting when
performed on the eBOSS LRG and QSO datasets and discuss the systematics possible physical origins. Finally, we conclude in \S\ref{sec:conclusion}.

In this work, all theoretical calculations and redshift-to-distance conversion assume a best-fit Planck 2018 $\Lambda$CDM cosmology \citep{PlanckCollaboration:2020A&A...641A...6P}.

\section{Spherical Fourier-Bessel Formalism}
\label{sec:sfb_basis}
The SFB basis is composed of eigenfunctions of the Laplacian in spherical coordinates, namely the product of the angular basis functions, spherical harmonics $Y^*_{\ell m}(\hat{\bx})$, and the radial basis functions $g_{n\ell}(x)$. The radial functions are the linear combinations of the spherical Bessel functions of the first kind and second kind \citep{Samushia:2019}:
\begin{align}
g_{n\ell}(x) = c_{n\ell} \, j_\ell(k_{n\ell}x) + d_{n\ell} y_\ell(k_{n\ell}x)\,,
\label{eq:gnl_basis}
\end{align}
where $k_{nl}$ denotes the total Fourier magnitude (the eigenvalue of the Laplacian) of each SFB mode.  The SFB modes are properly discretized under a spherical shell with finite comoving distance range $x_{\rm min}\leq x \leq x_{\rm max}$, with $n$ denoting the index for wave number $k_{n\ell}$ at each angular multipole $\ell$. The constants $c_{n\ell}$ and
$d_{n\ell}$ are chosen to satisfy the orthonormality relation
\begin{align}
\label{eq:gnl_orthonormality}
\int_{x_{\rm min}}^{x_{\rm max}}\dd{x}\,x^2g_{n\ell}(x)g_{n'\ell}(x)
&=
\delta^{\rm K}_{nn'}\,.
\end{align}

Beyond discretizing the SFB modes, the index $n$ characterizes the number of zero-crossings in the radial basis functions, corresponding to the number of half-cycles in their radial oscillations \citep{Wen+:2025PhRvD.112}. As such, increasing $n$ captures finer structure along the radial direction. The separation between angular and
radial scales is then achieved through the two indices $\ell$ and $n$ in the SFB basis.

In this section, we describe the estimator and theory for SFB power spectrum. We emphasize that the formalism has been validated on eBOSS EZmocks \citep{GrasshornGebhardt+:2024PhRvD.109h3502G}, and it is sufficiently accurate for the statistical precision of eBOSS.

\subsection{Estimator}
We use the public code \href{https://github.com/hsgg/SphericalFourierBesselDecompositions.jl}{\texttt{SuperFaB}} for estimating SFB power spectrum (PS). We refer the
interested reader to \citet{GrasshornGebhardt+:2021PhRvD.104l3548G} and give only an overview here. For a galaxy catalog, the density fluctuation can be measured as
\ba
  \delta^\textrm{obs}(\bx)
    &= w(\bx)\,\frac{n_{\rm g}(\bx) - \alpha\,n_{\rm s}(\bx)}{\nbar_\max}\,.
  \label{eq:delta_fkp_weighted}
\ea
Here $n_{\rm g}(\bx)=W(\bx)n(\bx)$ is the observed galaxy number density under the window function $W(\bx)$, $n_{\rm s}(\bx)$ is the number density for a synthetic random catalog, and $\alpha$ is the ratio of the
number of observed galaxies in the catalog to the number of objects
in the synthetic catalog. $\bar{n}_{\rm max}$ is the maximum galaxy number density in the data catalog, and the window function is normalized to unity at the maximum number density in our convention.

We use the FKP weight $w(\bx)$ to improve the statistical power of power spectrum measurements \citep{Feldman+:1994ApJ...426...23F}
\ba
\label{eq:fkp_weights}
w(\bx) &= \frac{1}{1 + n_{\rm g}(x) \, C_{\ell nn'}}\,,
\ea
where we approximate $C_{\ell nn'}\sim3\times\SI{e4}{\per\h\cubed\mega\parsec\cubed}$ as a constant, and $n_g(x)$ is the observed radial number density of the galaxies. Optimal weights in SFB may take a different form than those proposed for the Cartesian Fourier-space power spectrum by \citet{Castorina+:2019JCAP...09..010C}, and we defer a study of this to future work.

The synthetic random catalog is generated such that it describes the expected density of observed galaxies under the survey geometry and selection in the absence of cosmological clustering
\begin{align}
    \alpha n_{\rm s}(\bx)=W(\bx)\bar{n}(x)\,,
    \label{eq:synthetic-catalog}
\end{align}
where $\bar{n}(x)$ is the estimated mean number density based on the data catalog
\begin{align}
\bar{n}(x)&=\frac{1}{\int d\hat{\bx}\, W(\bx)}\int d\hat{\bx}\, W(\bx)n(\bx)\,.
\label{eq:radial-IC-mean}
\end{align}
Due to the difficulty of measuring radial selection function of galaxy surveys, we enforce the observed galaxy redshift distribution directly onto the synthetic random catalog (\cref{eq:synthetic-catalog}). The effect of this procedure on the clustering statistics is known as the radial integral constraint \citep{deMattia+:2019JCAP...08..036D}, which we model in \ref{sec:geometric}.

We can then perform the SFB transform as follows
\begin{align}
\hat{\delta}_{n \ell m}^{\textrm{obs}}&= \int \dd^3\bx\,g_{n\ell}(x)\,Y^*_{\ell m}(\hat{\bx})\delta^{\textrm{obs}}(\bx)\,,
\label{eq:sfb_discrete_fourier_pair_b}
\end{align} 
where the transform occurs over the spherical shell of $x_{\rm min}\leq x \leq x_{\rm max}$. We first perform the radial integral for each $(n,\ell)$ combination by directly
summing over the galaxies \citep{Leistedt+:2011ascl.soft11011L} and pixelize on the spherical sky using the HEALPix scheme \citep{Gorski+:2005ApJ...622..759G}. The angular integration is
then performed using \texttt{HEALPix.jl} \citep{Tomasi+:2021ascl.soft09028T}. This SFB transform is performed for both the data and the random
catalogs. 

Finally, we construct the pseudo-SFB power spectrum
\ba
\hat C^{\textrm{obs}}_{\ell nn'}
&= \frac{1}{2\ell+1} \sum_{m} \hat\delta^{\textrm{obs}}_{n\ell m} \hat\delta^{\textrm{obs}*}_{n'\ell m}\,.
\label{eq:pseudo-SFB-PS}
\ea

\subsection{Power spectrum theory}
\label{sec:power_spectrum_theory}

Here we give an overview on the computation of the SFB power spectrum under the linear perturbation theory. We refer interested readers to \citet{GrasshornGebhardt+:2021PhRvD.104l3548G} and \citet{Wen+:2024PhRvD.110l3501W} for theoretical and numerical details.

In the absence of window function, integral constraints, and shot noise, the theoretical SFB PS is
\begin{align}
  \langle\delta_{n \ell m}\delta_{n' \ell' m'}\rangle =\delta_{\ell\ell'}^{\rm K}\delta_{mm'}^{\rm K}C_{\ell n n'}\,.
\end{align}
The translational invariance is broken in the presence of redshift evolution and redshift space distortions (RSD), while the rotational invariance is still preserved. This leads to the angular mode $\ell$ and two radial modes $n,n'$ (one for each argument of the two-point function) as a complete decomposition of the two-point statistics for the overdensity field.

Under linear Newtonian RSD, we can calculate the SFB power spectrum as \citep{GrasshornGebhardt+:2021PhRvD.104l3548G}
\ba
C_{\ell nn'}
&=
\int_0^\infty\dd q\,\mathcal{W}_{n\ell}(q)\,\mathcal{W}_{n'\ell}(q)\,P_{m,0}(q)\,,
\label{eq:SFB-theory}
\ea
where $P_{m,0}(q)$ is the linear matter power spectrum at the present time, and $q$ denotes the Fourier mode of the underlying matter field. The SFB kernel $\mathcal{W}_{n\ell}(q)$ is given by
\ba
\label{eq:Wnlq}
\mathcal{W}_{n\ell}(q)
&=
\sqrt{\frac{2}{\pi}}\,q
\int_{x_\min}^{x_\max}\dd x\,x^2\,g_{n\ell}(x)\,D(x)
\vs&\times
e^{\frac12\sigma_u^2q^2\partial_{qx}^2}
\left[
  b(x,q)\,j_\ell(qx) - f(x)\,j''_\ell(qx)
\right],
\ea
where $g_{n\ell}(x)$ is the SFB radial basis functions defined in \cref{eq:gnl_basis}, $D$ is the linear growth factor, and $f$ is the linear growth rate.

The exponential operator in \cref{eq:Wnlq} describes the nonlinear Gaussian velocity dispersion at small scales. We approximate the Gaussian damping by expanding the exponential
operator (that acts on the spherical Bessels only) in a Taylor series to get \citep{GrasshornGebhardt+:2024PhRvD.109h3502G}
\ba
e^{\frac12\sigma_u^2q^2\partial_{qx}^2} j^{(d)}_\ell(qx)
&\approx
j^{(d)}_\ell(qx)
+ \frac{q^2}{2}\sigma_u^2(x)\, j^{(d+2)}_\ell(qx)\nonumber\,,
\ea
where $j_\ell^{(d)}(qx)$ is the $d$th derivative with respect to $qx$. The velocity dispersion $\sigma_u(x)$ is a combination of the velocity dispersion for Fingers-of-God (FoG) effect and redshift measurement uncertainty.

Local non-Gaussianity is modeled via a scale-dependent bias
\citep{Dalal+:2008PhRvD..77l3514D}
\ba
b(x,q) &= b_1(x) + \fnl b_\phi(x)\,\frac{3\Omega_m H_0^2}{2 q^2 T(q) \bar D(x)}\,,
\label{eq:bias}
\ea
where $b_1$ is the linear galaxy bias, and $b_\phi$ is the local PNG bias given the response to the presence of local PNG of the tracer. Here $H_0$ is the present Hubble parameter, $T(q)$ is the matter transfer function, and $\bar D(x)$ is the growth factor normalized to the scale factor $a$ during matter domination $\bar D(x)= (1 + z_\mathrm{md})^{-1}D(x)/D(x_\mathrm{md})$, where the $x_{\rm md}$ indicates a time deep within matter domination.

The theoretical prescription of the local PNG bias $b_{\Phi}$ has been under extensive discussion in recent literature\footnote{For recent examples, see \citet{Hadzhiyska+:2025,Sullivan+:2025,Dalal+:2025,Shiveshwarkar:2025+,Perez+:2026,Moore+:2026} and the references therein.}. Here we simply follow \citet{Slosar+:2008JCAP...08..031S} and assume the following relation
\ba
b_\phi(x) &= 2\delta_c(b_1(x) - p)\label{eq:universal-bias}\,,
\ea
where $\delta_{\rm c}=1.686$ is the critical density for spherical collapse and $p$ quantifies the merger history of the tracer. Such forms of LPNG bias have been deployed in most PNG analyses to date. We use the universality relation ($p = 1$), that is assuming that their halo occupation distribution (HOD) depends only on halo mass \citep{Tellarini+:2015}, as the default for our analyses. This is consistent with the previous LRG analysis \citep{Chaussidon+:2025JCAP...06..029C}. For the QSO sample, it is more customary in the literature to use $p=1.6$ \citep{Chaussidon+:2025JCAP...06..029C,Castorina+:2019JCAP...09..010C,Cagliari+:2024JCAP...08..036C}, assuming that all the quasars have a recent merger history \citep{Slosar+:2008JCAP...08..031S}.
We investigate the impacts of switching to $p=1.6$ for our analyses in \cref{sec:change-p}.

To evaluate the SFB PS, the bulk of the computation is spent on the spherical Bessel functions in the SFB kernel $\mathcal{W}_{n\ell}(q)$ of \cref{eq:Wnlq}. However, the Bessel functions only depend on the
combination $qx$, not on $q$ and $x$ separately. Thus, we can choose
discretizations for $q$ and $x$ such that in $q$-$x$ space the ``iso-$qx$'' lines go precisely through grid points. This Iso-qr integration method was proposed in \citet{GrasshornGebhardt+:2024PhRvD.109h3502G} and explained in detail in \citet{Wen+:2024PhRvD.110l3501W}, achieving evaluation of the SFB PS under linear Newtonian RSD on the order of seconds on a single CPU and enabling inference on data.

\subsubsection{Chebyshev expansion}
\label{sec:sfb_formalism}
\begin{figure}
    \centering
    \incgraph{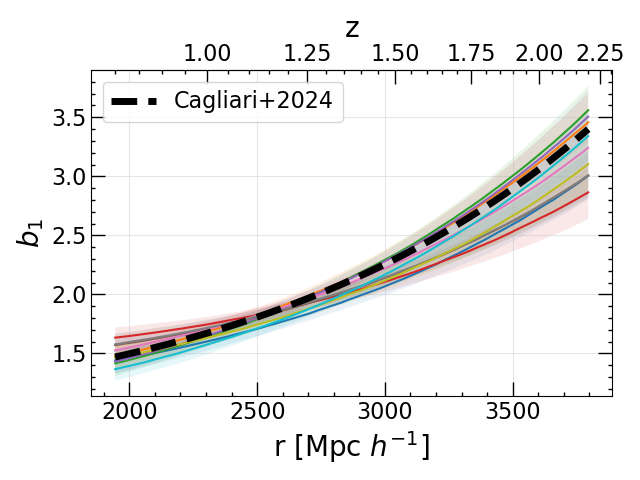}
    \caption{The fit QSO bias for 10 EZMocks (colored lines, 16th and 84th quantiles shaded) as a function of radial distance. The upper axis shows the equivalent $z$. We overplot the fit from \citet{Cagliari+:2024JCAP...08..036C} in black. There is good agreement between our EZMock fits and the result from \citet{Cagliari+:2024JCAP...08..036C}, validating our Chebyshev based bias model.}
    \label{fig:bias_model}
\end{figure}

The linear galaxy bias $b_1$ is redshift-dependent for selected galaxy samples, and here we use the Chebyshev polynomials $T_n$ up to the $N{\rm th}$ order to model its redshift evolution
\ba
b_1(z) &= \frac{\sum_{n=0}^{N}b_{1,n}T_n(\tilde{x}(z))}{D(z)}\,.
\ea 
We performed the Chebyshev expansion over the normalized comoving distance $\tilde{x}(x(z)) = (x(z) - x_\mathrm{mid})/\Delta x$ where $x_\mathrm{mid}=\frac12(x_\max+x_\min)$ and $\Delta x=\frac12(x_\max-x_\min)$. The Chebyshev coefficients $b_{1,n}$ will be the fitting parameters. In practice, we find two coefficients provide a sufficient description for the bias evolution of both LRG and QSO samples, that is
\ba
\label{eq:bias_chebyshev}
b_1(z) &= \frac{b_{1,0} + b_{1,1} \, \tilde{x}(z)}{D(z)}\,,
\ea
a linear function with respect to the comoving distance modulated by the growth factor. We tested using a quadratic function instead of the linear function and found that was not needed to emulate the bias's redshift evolution. Figure \ref{fig:bias_model} shows ten QSO EZmocks fit biases values against values from \citet{Cagliari+:2024JCAP...08..036C}, showing good agreement.

For the velocity dispersion including FoG and redshift errors, we also use Chebyshev polynomials to model the modulation over some fiducial velocity dispersions
\begin{align}
    \label{eq:sig_fogz_expansion}
    \sigma_u(x)=\sigma_u^{\rm ref}(x)\left(\sum_{n=0}^{N}\sigma_{u,n}T_n(\tilde{x}(x))\right)\,,
\end{align}
where the Chebyshev coefficients  $\sigma_{u,n}$ allow marginalization over the dispersion uncertainties. For LRGs, we follow \citet{GrasshornGebhardt+:2024PhRvD.109h3502G} and use $\sigma_u^{\rm ref}(x)^2=f(x)^2\sigma_{u,{\rm fog}}^2+\sigma_{u,{\rm z}}^2$ that explicitly combines the FoG and redshift error components. The redshift measurement uncertainty is kept constant \SI{1.05}{\per\h\mega\parsec} across the LRG redshift range \citep{Ross+:2020MNRAS.498.2354R}, while the FoG dispersion $\sigma_{u,{\rm fog}}$ is set from the best fit to LRG EZmocks.

For the QSO sample we adopt a simpler modeling scheme with the fiducial function set to $\sigma_u^{\rm ref}=1$, letting the Chebyshev decomposition fully capture the redshift evolution. This is motivated by our experience with the LRG sample and explicit checks on the impacts of velocity dispersion, finding that the FoG evolution is simple enough to be captured by this model. In the default analysis, we fix LRG to its fiducial function, and set QSO only through the first Chebyshev coefficient with $\sigma_{u,0}=5.5\,h^{-1}{\rm Mpc}$, obtained by fitting the QSO EZmocks, without marginalizing over the dispersion uncertainties. Due to the large redshift span and larger redshift uncertainties of QSO, we explicitly show the impact of marginalizing over the dispersion parameters for QSO inference in \cref{sec:fog-test}, demonstrating that the dispersion modeling does not impact our results.

\subsection{Estimator effects}\label{sec:geometric}
In this work, we forward model the observational effects from the estimator, including window function, radial integral constraint, shot noise, and pixel windows, when compared to the SFB PS measurements. The ensemble average of the pseudo-SFB PS estimator (\cref{eq:pseudo-SFB-PS}) is related to the theoretical SFB PS (\cref{eq:SFB-theory}) as the following
\begin{align}
 C_{\ell n_1 n_2}^{\textrm{obs}}&\equiv \langle \hat{C}_{\ell n_1 n_2}^{\textrm{obs}} \rangle\nonumber\\
 &=N^{\textrm{obs}}_{\ell n_1 n_2}+B_{\ell}\sum_{L N_1 N_2} \mathcal{A}^{L N_1 N_2}_{\ell n_1n_2} C_{L N_1 N_2}\,,
 \label{eq:SFBPS-IC}
\end{align}
where $N^{\textrm{obs}}_{\ell n_1 n_2}$ is the shot noise, and $B_{\ell}$ is the angular HEALPIX pixel window function. There are no pixel window effects in the radial direction since the radial transform is done exactly without binning \citep{Leistedt+:2012A&A...540A..60L,GrasshornGebhardt+:2021PhRvD.104l3548G}.

The PS mixing matrix $\mathcal{A}$ depends on the window (including weight) matrix $wW$ and the integral constraint matrix $wG$ \citep{Wen+:2026}
\begin{align}
  \mathcal{A}&=\mathcal{M}[wW-wG,wW-wG]\label{eq:A-mixing}\,.
\end{align}
Here we use the notation $\mathcal{M}[A,B]$ to denote the azimuthally averaged contraction between two SFB matrices $A$ and $B$
\begin{align}
\mathcal{M}[A,B]_{\ell n_1 n_2}^{L N_1 N_2}=\frac{1}{2\ell+1}\sum_{m,M} A_{n_1\ell m}^{\phantom{xx}N_1LM}B^{n_2\ell m}_{\phantom{xx}N_2LM}\,.
\label{cref:pCl-mixing}
\end{align}
The window mixing matrix (including weight) in the SFB space is a double SFB transform of $w(\bx)W(\bx)$
\begin{align}
(wW)_{n_{\rho}\ell_{\rho}m_{\rho}}^{\phantom{xx}n_{\mu}\ell_{\mu}m_{\mu}}&=\int d^3\bx\, g_{n_{\rho}\ell_{\rho}}(x)Y_{\ell_{\rho}m_{\rho}}^{*}(\hat{\bx})\nonumber\\
&g_{n_{\mu}\ell_{\mu}}(x)Y_{\ell_{\mu}m_{\mu}}(\hat{\bx})w(\bx)W(\bx)\label{eq:double-window-transform}\,,
\end{align}
and the mixing matrix for radial integral constraints are
\begin{align}(wG)_{n_{\rho}\ell_{\rho}m_{\rho}}^{\phantom{xx}n_{\mu}\ell_{\mu}m_{\mu}}=4\pi\sum_{n'}(wW)_{n_\rho\ell_\rho m_\rho}^{\phantom{xx}n'00}\widetilde{W}^{\phantom{xx}n_{\mu}\ell_{\mu}m_{\mu}}_{n'00}\,,
     \label{eq:RIC-SFB}
\end{align}
where $\tilde{W}$ is the radially normalized window function $\widetilde{W}(\bx)=W(\bx)/\int d\hat{\bx}\, W(\bx)$. Our transform and index notations for the window functions and integral constraints exactly follow Sec.~IV of \citet{Wen+:2026}. We refer readers to \citet{GrasshornGebhardt+:2024PhRvD.109h3502G} and \citet{Wen+:2026} for the details on computing these mixing matrices from random catalogs.
All above matrices arise from purely geometric effects for observations with no cosmological dependence.

\section{Inferring $\fnl$ from LRGs and QSOs}
\label{sec:fnl_inference}

\subsection{EBOSS }
\label{sec:eboss}
The data used herein are from the final extended Baryon Oscillation
Spectroscopic Survey (eBOSS) catalogs, constructed from the Sloan Digital Sky Survey (SDSS)-IV
Data Release 16 (DR16), described in detail in
\citet{Ross+:2020MNRAS.498.2354R}. The eBOSS DR16 LRG North Galactic Cap (NGC) sample contains
107,500 galaxies with spectroscopically measured redshifts over a range
$0.6 < z < 1.0$ and an angular extent of 2,566 deg$^2$, representing the combination of the eBOSS LRG sample and the overlapping portion of the
SDSS-III BOSS LRG sample \citep{Reid+:2016MNRAS.455.1553R}. The NGC QSO sample
contains 218,209 quasars over a redshift range $0.8 < z < 2.2$ and an area
2,924 deg$^2$. Data are also provided for the South Galactic Cap (SGC), though only those from the NGC are analyzed herein to avoid confounding systematics across the two fields, which may differ.

For each tracer and coverage region, random catalogs are provided alongside the
galaxy catalogs at a 50$\times$ higher density with randomly shuffled redshifts.
Three distinct sets of completeness weights are provided: ``close-pair'' weights
$w_{\text{cp}}$ which correct for the density modulation by spectroscopic fiber
collision; ``no z'' weights $w_{\text{noz}}$ which correct for the redshift
failure rate; and imaging systematics weights $w_{\text{sys}}$ which attempt to
correct systematic density modulations inherited from the associated imaging
survey. The total completeness weight is then taken as the product of these,
$w_{\text{c}} = w_{\text{cp}}w_{\text{noz}}w_{\text{sys}}$.

The associated EZmock mock catalogs \citep{Zhao+:2021MNRAS.503.1149Z} were
constructed using the effective Zel'dovich approximation and a standard flat
$\Lambda$CDM cosmology with $\Omega_{\text{m}} = 0.31$. Both ``complete'' and ``realistic'' catalogs
are provided, the former representing an ideal galaxy catalog observed
perfectly, free from any observational effects, and the latter representing
catalogs which have had all known observational and systematic effects applied
to both the data and random catalogs.

\subsection{Likelihood and Covariance Matrix}
\label{sec:mcmc}
In this section, we briefly describe our procedure for constructing the likelihood and obtaining parameter constraints. More details are provided in \citet{GrasshornGebhardt+:2024PhRvD.109h3502G}.

We use a Gaussian likelihood for the power spectrum,
\begin{align}
\label{eq:likelihood}
    \ln\mathcal{L} = \mathrm{const} - \frac12\,\Delta^T\,M^{-1}\,\Delta\,,
\end{align}
where $\Delta$ is the difference between the measured and theoretical SFB power spectra, and we estimate the covariance matrix $M$ from the \num{1000} realistic EZmocks.

However, since the number of modes in the SFB power spectrum is a factor of a few larger than \num{1000}, the covariance matrix obtained from these simulations has a large number of vanishing eigenvalues. In order to still use this estimate, we use the procedure described in \citet{Wang+:2020JCAP...10..022W} to derive an optimal compression of the data vector based on the nonzero eigenvectors of the covariance matrix. In brief, we perform an eigendecomposition of the full covariance matrix, then we assemble a compression matrix $R$ from the eigenvectors which, following \citet{GrasshornGebhardt+:2024PhRvD.109h3502G}, have $\lambda_i > \frac{\lambda_{max}}{1000}$, where $\lambda$ are the eigenvalues. Our data vector and covariance matrix then become
\begin{align}
    \Delta &= R \cdot \big(\hat C_{\ell nn'}^{\mathrm{obs}} - C_{\ell nn'}^{\mathrm{theory}}\big)\,,\\
    M &= R \cdot \hat\Sigma \cdot R^T\,,
\end{align}
where $\hat\Sigma$ is the full covariance matrix estimated from the 1,000 EZmocks. The compression matrix $R$ projects the power spectrum onto the linear combination of modes that had a successful covariance estimate. The size of the uncompressed and compressed data vectors for the QSO and LRG samples 
depend on the mode cuts imposed, which we will vary in Section \ref{sec:cleaning}. For LRGs (QSOs), the uncompressed data vectors range in size from 6140 (13115) modes with only $\ell >8$ down to $\sim1500$ ($\sim2900$) modes after imposing $\ell$ bin cuts (see Section \ref{sec:binned_angular_scales}). The compressed data vectors then vary in size from 659 to 999 in LRGs, while all compressed QSO data vectors are of length 999.
\citet{2007A&A...464..399H} showed that inverting noisy covariance matrices can lead to overconfident (artificially small) confidence regions and presented a correction factor. We note here that we do not implement such a correction because our compression methodology diagonalizes $M$, a different regime from that considered by \citet{2007A&A...464..399H}. As we will show in Figure \ref{fig:ezmock_ell8_cut}, the inferred $\fnl$ values are unbiased at magnitudes relevant to the work here. Further, we can compare the EZmock-to-EZmock scatter on $\fnl$ and the mean of the error bars for the mocks. For the results shown in Figure \ref{fig:ezmock_ell8_cut}, the ratio $\langle \sigma \rangle / S$ = 0.95, where $\langle \sigma \rangle$ is the mean of the error bars on the inferred $\fnl$ across the mocks and $S$ is the standard deviation of the inferred $\fnl$ values across the mocks. For the bins considered in Section \ref{sec:cleaning}, this ratio has a mean of 1.09 and a standard deviation of 0.12, so the error bars may actually be marginally overestimated. The minimal coverage fraction is 0.95, which is consistent with 1 for the number of mocks we consider. We conclude that the compression algorithm does not meaningfully bias $\fnl$ nor underestimate the error bars.

The likelihood in \cref{eq:likelihood} is sampled by an adaptive Metropolis-Hastings (AMH) sampler \citep{Roberts+:doi:10.1198/jcgs.2009.06134}, which we restart several times for faster convergence. The AMH sampler adjusts the proposal distribution every few steps by estimating a new proposal distribution from previous samples. Thus, it is Markovian asymptotically. We restart the sampler with an initial parameter vector nearer the maximum likelihood, and with a proposal distribution estimated from the previous run. This helps with achieving the asymptotic state of the AMH quicker.

The full set of parameters considered in the inference in the main text is $\{b_{1,0}, b_{1,1}, \fnl\}$, where  $b_{1,i}$ parameterize the linear galaxy bias in \cref{eq:bias_chebyshev}. We always marginalize over the two linear bias parameters in inference. In \cref{sec:fog-test}, we additionally marginalize over $\{\sigma_{u,i}\}$, which parametrize the velocity dispersion in \cref{eq:sig_fogz_expansion}.  

\subsection{Baseline $\fnl$ Inference}
\label{sec:inital_inference}

With the eBOSS LRG and QSO datasets, SFB estimator, likelihood, and covariance in hand, we can perform inference on the set of  parameters. 

As an initial set of minimal cuts, we impose that $\ell_{min}$ is 8 -- this removes the largest of the angular scales that exceed the limited footprint of the survey. As can be seen by comparing the upper- and lower-left panels in Figure \ref{fig:sfb_demo}, removing the $\ell < 8$ modes cuts modes which are rapidly approaching zero due to the integral constraint.
We impose that $0.0 < k < 0.08\; h/\mathrm{Mpc} $ which excludes small scales where nonlinearity becomes non-negligible. Finally, we choose that $\Delta n < 2$, where $\Delta n = (n-n')$ which removes modes beyond the first two off-diagonals of $C_{\ell n n\prime}$; for $\Delta n > 2$, the power spectrum rapidly approaches zero.  

\begin{figure}
  \centering
  \incgraph{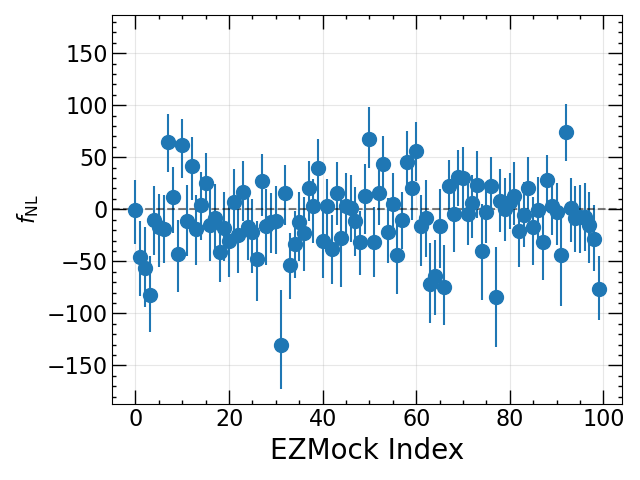}
  \caption{The inferred median $\fnl$ values for 100 LRG EZMocks with $\ell_{min} = 8$ when fitting for $\fnl$ and the bias parameters. The error bars are the 16th and 84th quantile of the posterior. The median value across all 100 EZmocks is $\fnl = -8$ with $\frac{\sigma_{\fnl}}{\sqrt{N}} = 3.4$. The estimator is slightly negatively biased, but at a magnitude much smaller than the variances in $\fnl$ that we will attempt to probe in this paper.}
  \label{fig:ezmock_ell8_cut}
\end{figure}

\citet{GrasshornGebhardt+:2024PhRvD.109h3502G} showed that the estimator, in the ensemble, recovers an $\fnl=0$ signal in mocks. We would like to extend this conclusion to our set of cuts. To do so, we make an initial inference on $\fnl$, $b_{1,0}$, and $b_{1,1}$ from 100 EZmocks to quantify any bias in the posterior of $\fnl$ and quantify the expected scatter. Note that the compressed covariance ($M$, \ref{sec:mcmc}), which is used in the likelihood calculation for the AMH sampler, is recalculated after cutting the modes; in fact, the covariance is recalculated each time we make a unique set of mode cuts throughout the paper. The distribution of $\fnl$ from the EZmocks is shown in Figure \ref{fig:ezmock_ell8_cut}. We can see in Figure \ref{fig:ezmock_ell8_cut} that the inferred $\fnl$ value is consistent with zero at about the expected rate. The median recovered $\fnl$ value is $-8 \pm 3.4$, so the estimator is marginally negatively biased but at a magnitude that will prove to be insignificant for this paper. 

\begin{figure}
  \centering
  \incgraph{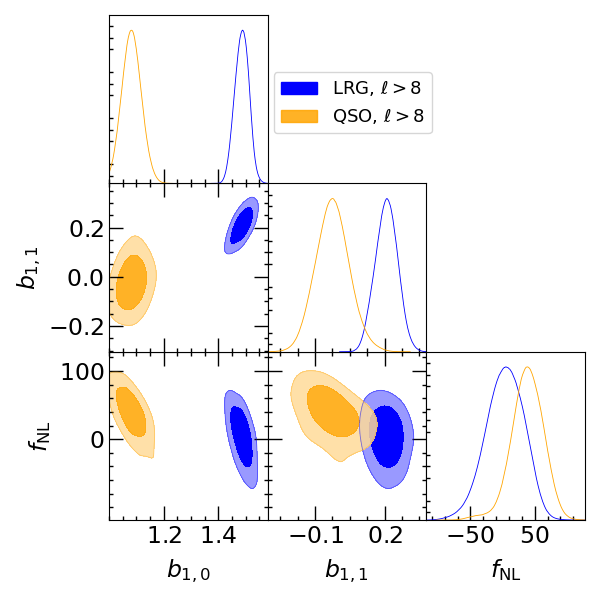}
  \caption{The posterior distributions on the bias parameters and $\fnl$ for the LRG and QSO samples with $\ell_{min} = 8$. The LRGs $\fnl$ posterior is zero centered at $\fnl=5^{+30}_{-29}$ while the QSOs $\fnl$ posterior is in slight tension with zero at $\fnl=38^{+25}_{-23}$. 
  }
  \label{fig:initial_inference}
\end{figure}

We then make the same selections on the LRG and QSO data vectors and do the inference on $\fnl$, $b_{1,0}$, and $b_{1,1}$. Figure \ref{fig:initial_inference} shows the resulting corner plots from these inferences. The LRG posterior on $\fnl$ has a median and 16th and 84th quantiles at $\fnl = 5^{+30}_{-29}$ (we will use the 16th and 84th quantiles as error bars on the posterior distribution of $\fnl$ throughout the paper). This constraint is consistent with values derived from Planck \citep{PlanckCollaboration:2020A&A...641A...9P} and LSS surveys \citep{Leistedt+:2014PhRvL.113v1301L, Castorina+:2019JCAP...09..010C}. The QSOs inferred $\fnl$ value, however, is in slight tension with $\fnl \simeq 0$, with $\fnl = 38^{+25}_{-23}$.

\section{Consistency Checks}
\label{sec:cleaning}
With the SFB basis we can push deeper into the inferred  $\fnl$ values. One may expect that systematics in the data may be (at least fairly) localized to particular angular or radial modes--for example, the characteristic angular size of the scan strategy of the survey or a redshift solution that gets particularly favored by the interference of a sky emission line that is not fully mitigated. We will now make a variety of cuts on SFB modes, rerun the inference, and check for self consistency on the inferred cosmological parameters, finding hints of residual systematics along the way.

The core idea behind our systematics identification procedure is that if one measures an SFB power spectrum data vector where all response is sourced by true cosmological signal, there should be consistency between the inferred values of $\fnl$ when removing angular and/or radial modes to within the error bars at the expected rates from Gaussian statistics. Thus, the fundamental workflow for our systematic hunt exploits the consistency of inference (or lack thereof) when dropping modes and redoing the inference. If the data were free from systematics, dropping modes would decrease the constraining power on $\fnl$ and the other fit parameters, but the new inferred value would be consistent with the old. This would manifest as inferred $\fnl$ values fluctuating about the original value with increased error bars on the parameter, but the value would be consistent. If, on the other hand, there are systematics present in the data, dropping the modes afflicted by systematics may result in the inferred value of $\fnl$ changing drastically, moving to a new value far outside the original error bars--further, the magnitude of the error bars themselves may change rapidly.

Note that genuine large-scale signals arising from linear relativistic effects \citep{Challinor+:2011PhRvD..84d3516C,Bonvin+:2011PhRvD..84f3505B} could also result in $\fnl$ values moving away from zero, since we ignore these terms in our Newtonian RSD modeling of \cref{eq:Wnlq}. However, the magnitudes of these shifts by ignoring the relativistic effects are generally $\Delta f_{\rm NL}\lesssim 5$ \citep{Foglieni+2023,Wen+:2026}, drastically smaller than the statistical sensitivities of the eBOSS samples and the shifts we will see throughout this section of the paper.

\begin{figure*}
  \centering
  \incgraph[]{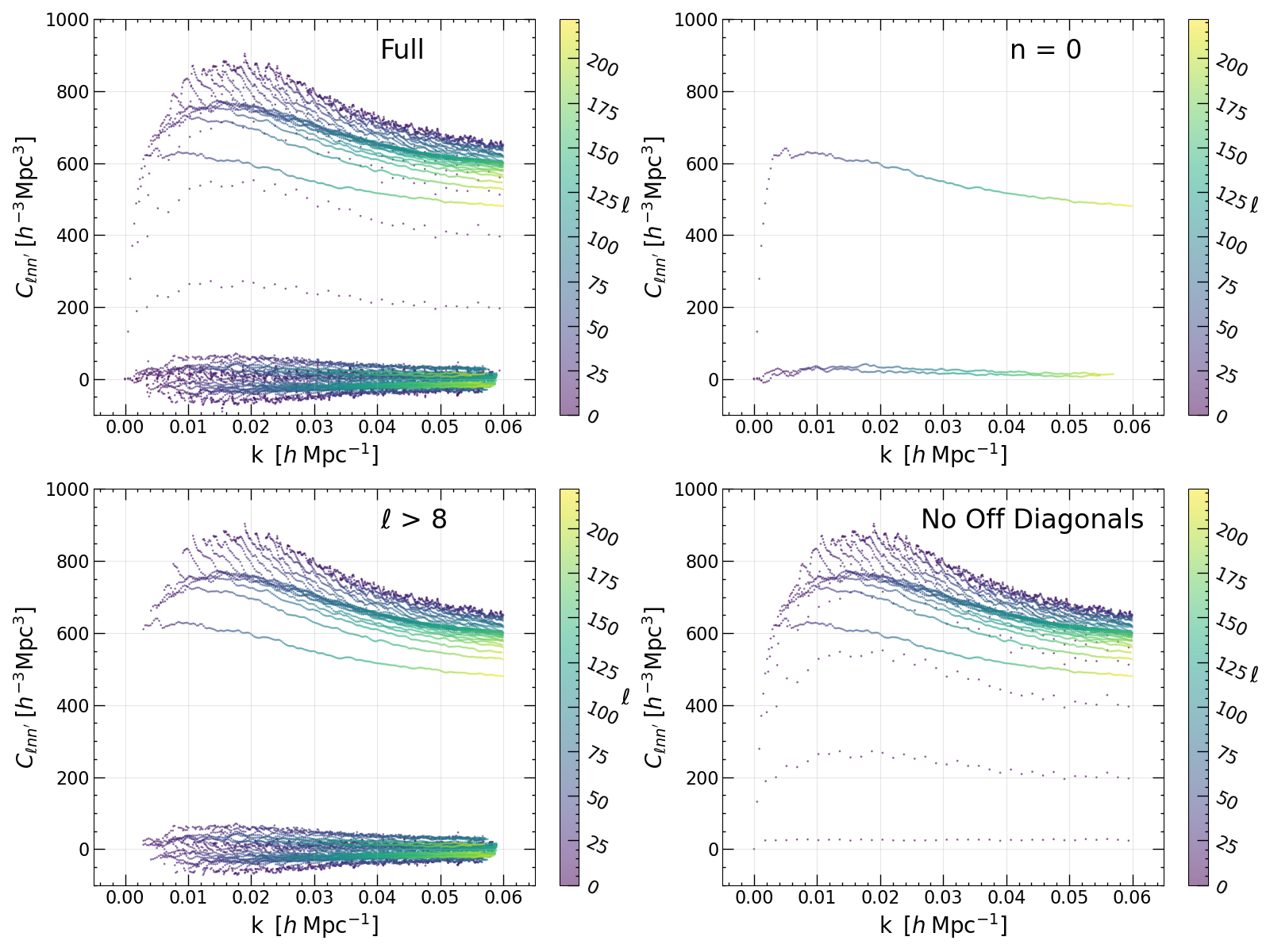}
  \caption{Various cuts on the average measured SFB power spectrum for 1,000 QSO EZmocks. Top left: the entire SFB power spectrum. The ranges of $n$ and $\ell$ considered can be determined by the chosen limits in $k$, in this case spanning $0 < n < 35$ and $0 < \ell < 222$. All points are color coded by $\ell$ for all plots on a set color scale. Top right: the $n=0$ mode with two off-diagonals, $n'=1$ and $n'=2$. The two off-diagonals are at the bottom of the plot. Continuous traces in the full power spectrum correspond to fixed $n$. Bottom left: The SFB power spectrum with $\ell_{min}$ = 8. The sharp downturns in the traces for all $n$ are removed. Bottom right: Removing the off-diagonals, which are near zero. The remaining points near zero are the smallest $\ell$ modes for all $n$. 
  }\label{fig:sfb_demo}
  
\end{figure*}

It is useful to look at plots of the SFB power spectrum for the EZMocks with some of the cuts that we will consider. Figure \ref{fig:sfb_demo} shows the average of 1,000 measured QSO EZMocks' SFB power spectra in four panels: the full data vector in the top left, the n=0 mode only in the top right, the $\ell > 8$ modes in the bottom left, and only the on-diagonals ($n=n'$) in the bottom right. We will explore the impact of making cuts of these types on the data vector to test how the inferred value of $\fnl$ changes. Some key things to note and keep in mind:
\begin{enumerate}
    \item A given $n$ mode expresses as a continuous ``trace'' in the SFB power spectrum with $\ell$ increasing as effective $k$ increases (the upper right panel shows the n=0 mode and its first two off-diagonals),
    \item low $\ell$ modes tend to turn over and rapidly approach zero power across $n$ modes (compare the modes removed between the upper- and lower-left panels), a result of the integral constraint, and 
    \item off diagonals ($n\neq n'$) tend to cluster near zero, a reflection of cosmological homogeneity (lower right panel) \citep{Khek+:2024PhRvD.110f3524K}. 
\end{enumerate}

\subsection{Radial Cuts}
\label{sec:radial_cuts}
\begin{figure}
  \centering
  \incgraph{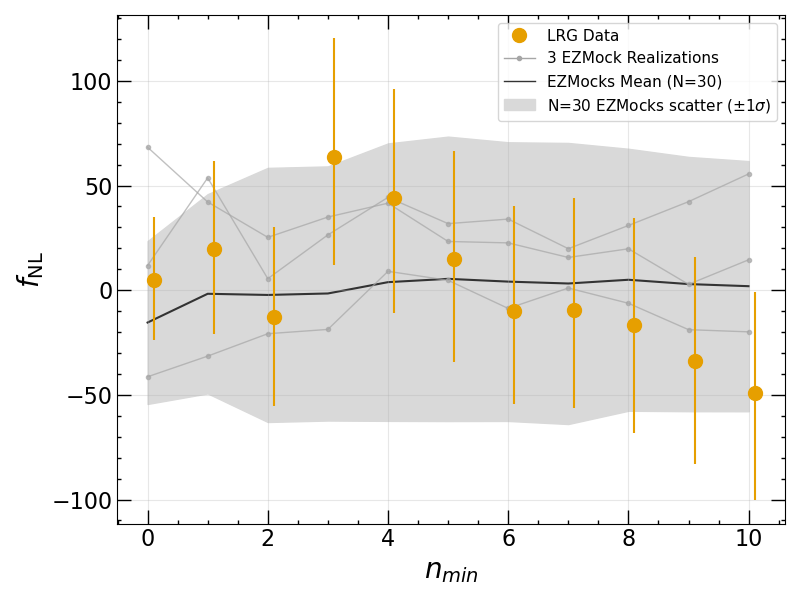}
  \caption{Inferred $\fnl$ as a function of $n_{min}$ for the LRG data (orange). Three individual EZmocks with the same successive cuts are plotted as gray traces, the mean of the $N=30$ mocks in black, and the 1-sigma scatter from the EZMocks is shown as a gray band. The data shows behavior that is qualitatively consistent with the mocks.}
  \label{fig:fnl_vs_nmin}
\end{figure}

We take 30 LRG EZMocks and successively increase the lowest $n$ kept in the data vector, $n_{min}$ and infer $\fnl$. Note that all of the cuts also include an $\ell_{min} = 8$ cut to remove modes which are too large to be probed by the survey footprint. Cutting the $n=0$ modes (i.e. imposing ${n_{min}}=1$) is equivalent to cutting power that varies only angularly, and is constant radially. Further, following the finding that stellar contamination tends to localize in the $n=0$ and low $\ell$ modes \citep{Wen+:2025PhRvD.112}, we set this $n_{min}$ threshold only for $\ell < 20$; in Figure \ref{fig:fnl_vs_nmin} we show the inferred $\fnl$ value for the progressively increasing $n_{min}$ where $\ell < 20$ for both the data and the EZMocks. The error bars on the data are the 16th and 84th quantiles of the posterior on $\fnl$.

Note that \citet{Wen+:2025PhRvD.112} find that the stellar contamination localizes to $n=0$, $\ell \lesssim 50$, but the threshold value of $\ell$ depends on the redshift of the sample, the contamination rate, and the contamination mechanism. Our results do not significantly change if we change this $\ell$ threshold. 

Figure \ref{fig:fnl_vs_nmin} shows no strong features in the data--there is a gradual rise and fall in the inferred $\fnl$ values that qualitatively appears to be consistent with its own error bars. Thus, in this test we do not find evidence for systematics localized to the radial modes in the LRG sample.

We will now consider the same exercise but varying $\ell_{min}$ instead of $n_{min}$.

\subsection{Angular Cuts}
\label{sec:angular_cuts}
\begin{figure}
  \centering
\incgraph{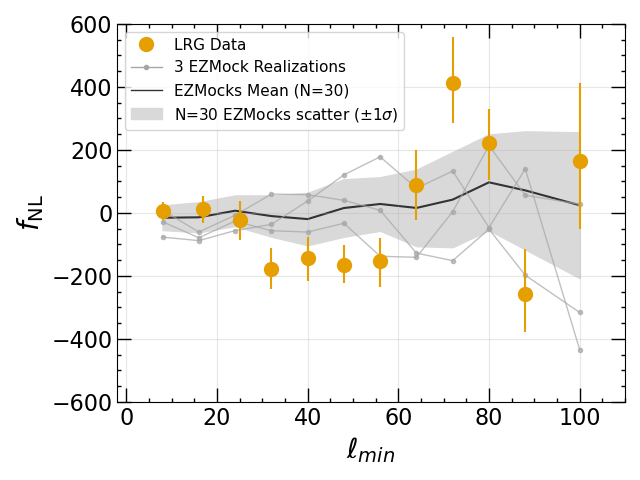}
  \caption{Inferred $\fnl$ as a function of $\ell_{min}$ for EZmocks (gray) and the LRG data (orange). We plot three example EZMocks to illustrate the covariance of adjacent $\ell_{min}$ data points. The EZMocks' ensemble mean is plotted in black and the 1-$\sigma$ scatter as a gray envelope. The LRG sample displays more sporadic behavior compared to the EZMocks, starting out near zero before falling to extremely negative values that are significantly far from zero when accounting for the error bars. This is an indication that there are unmitigated systematics within the data vector; the sudden sharp changes give an idea of where the systematics may be localized.}
  \label{fig:ezmock_vs_lrg_ellmin}
\end{figure}

Similar to before, we take 30 LRG EZMocks and successively increase the lowest $\ell$ kept in the data vector, $\ell_{min}$ (again starting with $\ell_{min}=8$), and infer $\fnl$. We repeat the same exercise for the actual data measurements for the eBOSS LRG data. The results are shown in Figure $\ref{fig:ezmock_vs_lrg_ellmin}$.

Three example EZMocks, in gray, with input $\fnl = 0$, yield inferred $\fnl$ values scattering about zero and, notably, with no sharp features as a function of $\ell_{min}$ until $\ell \gtrsim 75$. The covariance of adjacent data points explains the gradual rise and fall of inferred $\fnl$ across the three EZMocks.

The LRG data sample, on the other hand, displays sharp features where the inferred value of $\fnl$ changes rapidly as we increase $\ell_{min}$. 
The inference starts out at $\fnl \sim 0$ in the $8 \lesssim \ell_{min} \lesssim 24$ range, before rapidly decreasing to very negative $\fnl \sim -200$ values, and displaying more sporadic behavior at $\ell_{min} \gtrsim 60$. We can see that our initial inference, considered in Figure \ref{fig:initial_inference}, may have actually serendipitously been centered about $\fnl \sim 0$ as a result of systematics canceling each other out. The sporadic behavior in inferred $\fnl$ indicates that there are unaccounted for systematics in the sample and motivates cutting modes in the data vector to attempt to understand them better. 

In Figures \ref{fig:fnl_vs_nmin} and \ref{fig:ezmock_vs_lrg_ellmin}, we qualitatively looked for sharp features in the data that would indicate a systematic is present in either the angular or radial modes. Quantifying such features would be desirable but computationally expensive, and the physical interpretability of these figures is challenging. A more fruitful allocation of resources is to use these figures to motivate bin edges in a mode-binning scheme and to do many inferences on EZMocks and the data using these binned modes. With the EZMocks, this permits the quantification of tension between the data vector's inferred $\fnl$ and the expected value for an $\fnl \sim 0$ universe. Binning modes is far more physically intuitive than the $\ell_{min}$ and $n_{min}$ framework, since it allows us to identify parts of the data vector that appear to be afflicted by systematics, which can be directly mapped to angular scales. In this context, we investigate binning on the angular modes, with and without the $n=0$ mode, to push for a deeper understanding.

\subsection{Binned Angular Scales}
\label{sec:binned_angular_scales}
For the binned angular scales analysis, we consider not just the LRGs, but also the eBOSS QSO sample. We consider each sample independently, making the same cuts on both samples and inferring $\fnl$ from each sample separately. 

\begin{figure*}
  \centering
  \incgraph[]{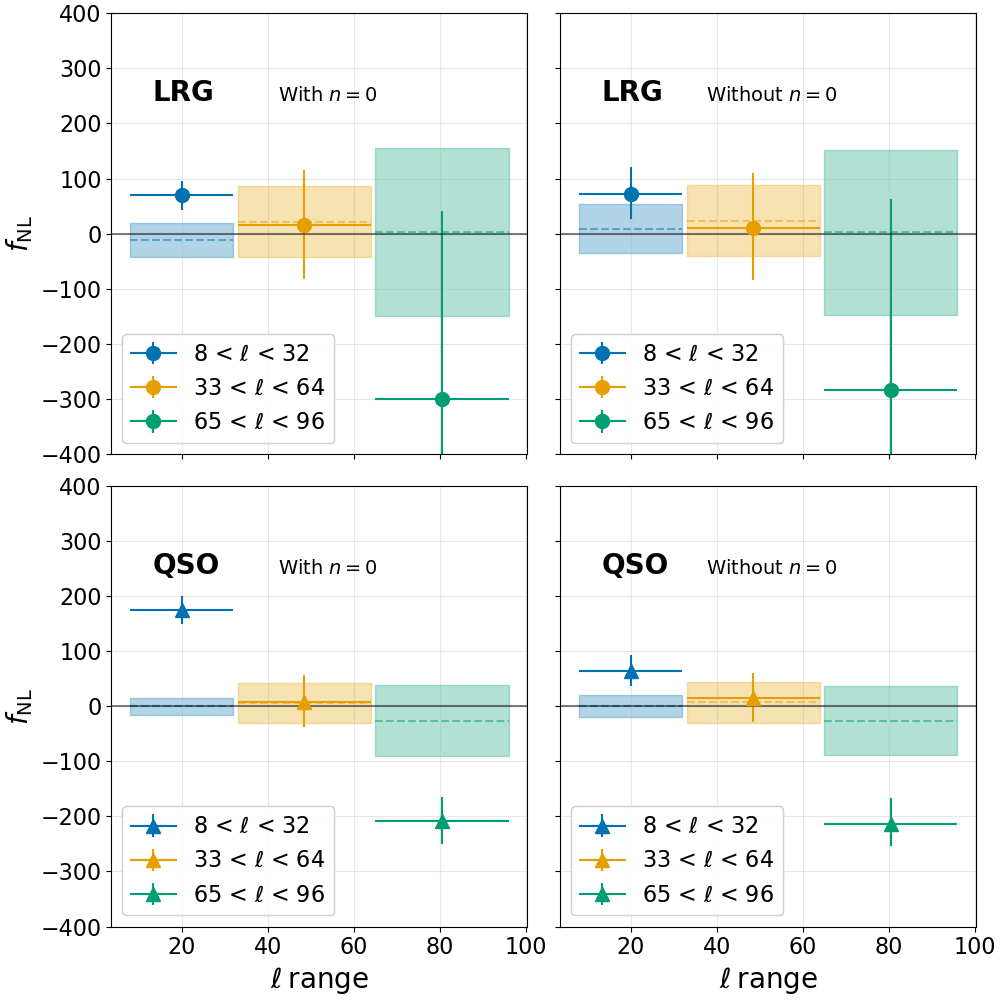}
  \caption{Inferred $\fnl$ for binned $\ell$ ranges: $8 \le \ell \le 32$, $33 \le \ell \le 64$, and $65 \le \ell \le 96$ in blue, orange, and green, respectively, with LRGs in the upper panels and QSOs in the lower panels. Left panels are with the $n=0$ mode and right panels are without it. The error bars are the 16th and 84th quantile from the posterior sampling. The horizontal dashed lines and shaded regions represent the mean and 1-sigma deviation of inferred $\fnl$ from a population of 50 EZmocks with the same cuts. The QSOs' smallest $\ell$ range, the blue point, shows signs of being afflicted by stellar clustering, which tends to impact the $n=0$ and low $\ell$ modes most strongly. Cutting the $n=0$ mode brings it into agreement with LRGs (though still in tension with the EZMocks). The largest $\ell$ bin, in green, shows a very strong negative bias for both samples, albeit with a large error bars for LRGs. This $\ell$ bin notably corresponds to the plate and imaging scale for eBOSS and its parent imaging survey. 
  }\label{fig:fnl_vs_binned_ell}
\end{figure*}

Motivated by the results in Figures \ref{fig:fnl_vs_nmin} and \ref{fig:ezmock_vs_lrg_ellmin}, we bin the $\ell$ modes for the LRG and QSO samples into ranges $8 \le \ell \le 32$, $33 \le \ell \le 64$, and $65 \le \ell \le 96$ and, to further investigate the impact of the $n=0$ mode, we make these splits with and without the $n=0$ mode included. We do the same splits across N=50 LRG and QSO EZMocks to quantify any discrepancies between the data and the EZMocks and to further stress test our modeling. The resulting inferred $\fnl$ values  for both the LRG and QSO samples are shown in Figure \ref{fig:fnl_vs_binned_ell}; the error bars on these points are the 16th and 84th quantiles from the posteriors. We show the LRGs in the upper panels and the QSOs in the lower panels; the left panels include the $n=0$ mode while the right panels excludes it. We fit the distribution of N=50 inferred $\fnl$ values from the EZMocks for each scenario with a Gaussian and plot the mean inferred value (dashed line) and the 1-sigma deviations as colored bands. In Appendix \ref{app:B}, we show the Gaussian fits to all EZMock populations in Figure \ref{fig:fnl_vs_binned_ell}. Since the distribution of EZMocks are consistent with Gaussians (see Appendix \ref{app:B}), we use a Student-t predictive distribution with $N-1=49$ degrees-of-freedom to calculate the p-values for the observed data points' tension with the EZMock distributions, accounting for having 50 fit mocks. These p-values are reported in Table \ref{tab:p-values}; a low p-value indicates that a systematic may be present in the modes considered.

\begin{table}[ht]
\caption {The Student-t predictive p-values of the data with respect to the distribution defined by the EZMocks for a given set of cuts.} 
    \centering
    \begin{tabular}{|c|c|c|c|}
     \hline
    & $8 \le \ell \le 32$ & $33 \le \ell \le 64$ & $65 \le \ell \le 96$  \\ \hline
    LRG w/ $n=0$ & 0.01 & 0.93 & 0.06  \\
    LRG w/out $n=0$ & 0.17 & 0.85 & 0.07  \\
    QSO w/ $n=0$ & <0.001  & 0.97 & 0.01  \\
    QSO w/out $n=0$ & 0.002 & 0.85 & 0.01 \\ 
    \hline
    \end{tabular}
    \label{tab:p-values}
\end{table}

There are many interesting features to digest in Figure \ref{fig:fnl_vs_binned_ell} and Table \ref{tab:p-values}. First, including the $n=0$ mode and considering the lowest $\ell$ range yields large positive $\fnl$ values, with the QSO sample's value farther from zero than the LRG sample; the p-value that the QSO data's $\fnl$ value was drawn from the distribution defined by QSO EZMocks is very close to zero. When we remove the $n=0$ mode, this tension is reduced owing almost entirely to a reduction in the offset from zero--the error bars stay approximately the same.

There are thus two provocative pieces of information: 1) the low $\ell$ range is biased high and this bias is worse for the QSO sample and 2) removing the $n=0$ mode from the QSO data vector removes a significant amount of this bias. 

Recalling the \citet{Wen+:2025PhRvD.112} result that stellar contamination largely resides in the $(n=0,\: \ell \lesssim 50)$ modes, one possible source of this systematic may be residual stellar contamination in the QSO sample. Foreground stars make identification of QSOs more difficult because of the increased background, which may imprint their clustering on to the QSO sample--the weights are meant to mitigate this, but may not do a perfect job. 

Another possibility comes from misidentification of QSOs and stars at the imaging level. The QSO target selection for eBOSS takes steps to remove stars from the proposed QSO sample--however, contamination at the imaging level is still expected, and although the contaminating stars will be nearly entirely removed from the QSO sample when their spectra are obtained, these stars may leave an imprint of stellar clustering on the QSO sample in two ways via fiber collisions.

Consider a true star and a true QSO near enough to each other to force a fiber collision, such that only one of these objects makes it into the QSO target sample. If the target preselection algorithm identifies both the star and the QSO as potential targets, the resulting fiber collision weight for whichever object is selected is increased. If the star is selected, targeted, identified to be a star, and discarded, we have missed a true QSO since there is no object in the QSO catalog to up-weight, as the star was rejected during the spectroscopy measurement. If the QSO is targeted, identified to be a QSO, and included in the sample, it is up-weighted to account for the fiber collision, though this weight was based on its collision with a star, not a QSO. Both of these failure modes imprint the stellar clustering onto the QSO clustering--the first through under-densities and the second through over-densities. In the auto power spectrum, both these under- and over-densities represent stellar-like clustering, which can contaminate large angular modes which are most sensitive to $\fnl$.

Also of note: as we discuss in more detail in   Appendix \ref{app:A}, the fit bias evolution for the QSO low $\ell$ bin is unphysical (it evolves in the opposite direction of the rest of the data bins and the EZMocks) and, because of its degeneracy with $\fnl$, this results in the inferred value of $\fnl$ being unstable. However, this behavior is unique to the data--we do not see equivalent instability in the fit bias evolution of the EZMocks when limiting to this $\ell$ range, indicating that the bias fit is being hampered by the systematic rather than by the limited $\ell$ range.

The middle $\ell$ bin gives an unbiased, though not particularly constraining inference on $\fnl$, a reflection of the fact that we have trimmed most of the modes and are not using a large data vector. However, its relative position compared to the blue and green points yields some insight about the features we saw in Figure \ref{fig:ezmock_vs_lrg_ellmin}, $\fnl$ as a function of $\ell_{min}$. In Figure \ref{fig:ezmock_vs_lrg_ellmin}, we saw that between $\ell_{min} = [8,24]$, we inferred $\fnl \simeq 0$. However, Figure \ref{fig:fnl_vs_binned_ell} shows that this was likely a serendipitous offsetting of different systematics that bias $\fnl$ in opposite directions. 

Finally, we consider the high $\ell$ bin. The LRG data points in both panels are biased toward very large negative $\fnl$ albeit with a large error bar that is consistent with zero. The QSO data points, on the other hand, have approximately the same offset from zero in $\fnl$ but with tighter error bars, putting them in contention with $\fnl \simeq 0$, the other $\ell$ bins, and the EZMocks; the p-values are both $0.01$ for the QSO inferred $fnl$ with respect to the EZMocks in this bin whether including the $n=0$ mode or not. The consistency between the LRG and QSO values in all $\ell$ bins, particularly when removing the $n=0$ mode, is both further validation of the estimator and evidence that the two samples share some unmitigated observer frame systematic, particularly in the lowest $8 \lesssim \ell \lesssim 32$. The QSOs appear to have a systematic affecting their $65 \lesssim \ell \lesssim 96$ bin which may be common with LRGs. It is noteworthy that this $65 < \ell < 96$ range roughly corresponds to the SDSS plate scale and the FoV of its parent imaging survey. In the course of attempting to isolate this systematic's source, we cut the off-diagonals ($n \neq  n'$), where inhomogeneity would localize in the SFB power spectrum, and found no improvement in the inferred $\fnl$ value. We also tested against model misspecification, changing $p=1.0$ to $p=1.6$ and allowing more redshift evolution in our FoG modeling. Both of these, however, also failed to explain the strong negative $\fnl$ bias. These adjustments and results are discussed in Appendix \ref{app:A}. Proper diagnosis of this systematic beyond the investigations in the Appendix would require efforts beyond the scope of this work, so we leave a deeper investigation to future researchers. 

\section{Conclusion}
\label{sec:conclusion}
In this work, we analyzed the clustering of the eBOSS LRG and QSO samples in the SFB framework. We sought to demonstrate finer mode cutting and systematic isolation in the SFB basis. The SFB basis allows one to naturally decompose systematics into angular and radial components on the sky, reflecting the natural geometry of astrometric surveys.

We operate under the assumption that the eBOSS LRG and QSO samples, if free of systematics, would lead to an inference of $\fnl \sim 0$ to within error bars, consistent with previous CMB and the latest LSS measurements. In this way, we are taking a decidedly \textit{unblinded} approach, using the inferred values of $\fnl$ to search for problematic regions of the data vector which may be afflicted by systematics. 

An initial inference of $\fnl$ from the LRG and QSOs samples yields values of $\fnl = 5 ^{+30}_{-29}$ and $\fnl = 38 ^{+25}_{-23}$, a perfectly reasonable inferred value from the LRGs and only a slight tension with expectations from QSOs. We use the SFB basis to dig deeper, searching for inconsistencies in the inferred value of $\fnl$ when we make particular mode cuts on the power spectrum, which would be an indicator of underlying systematics. We find that the $8 < \ell < 32$ bin, and within it particularly the $n=0$ modes (i.e., large-scale angular only modes) in the QSOs seem to be impacted by a systematic; under the assumption that the inferred $\fnl$ value is drawn from the same distribution as those from EZMocks with the same cuts applied, we calculate a p-value of $<0.001$. We conjecture that the systematic could be associated with stellar contamination, either through weights which do not fully mitigate their impact or the result of the stellar population imprinting onto the QSO target sample at the imaging level stage.

We also find that the QSO and LRGs samples have evidence ($p = 0.01$ for QSOs) of a systematic in the $65 < \ell < 96$ range, which roughly corresponds to the plate and imaging scales for eBOSS and its parent imaging survey. We attempt to isolate and cut this systematic in the off-diagonals, where radial inhomogeneities would arise, but do not see the inferred $\fnl$ value move toward zero after the cut. We rule out that the $p$ value or FoG modeling are causing this tension.

The analysis in this paper, while a proof of concept in using SFB for systematic identification, could not be replicated in an analysis where one is trying to infer $\fnl$ at a competitive level, where one cannot assume that $\fnl\sim 0$ at the level of the survey's sensitivity. Instead, the best practice would be to adopt a blinded approach. In a future blinded analysis, the hunt for unknown systematics could be accomplished by dropping random subsets of modes, checking for consistency in $\fnl$, and isolating modes which are common in the failing consistency checks.

\acknowledgments
The authors thank Alex Krolewski and Richard Feder for useful conversations that improved the paper's discussion of possible stellar contamination in the QSO sample and for giving feedback on a draft of this manuscript. The authors also thank {H\'ector Gil-Mar\'in} for useful discussion. The authors also thank the referee, who provided valuable feedback which increased the quality of this paper.
\textcopyright 2026. All rights reserved. Part of this work was done at Jet Propulsion Laboratory, California Institute of Technology, 
under a contract with the National Aeronautics and Space Administration. RYW acknowledges support through the Canada Graduate Research Scholarship-- Doctoral program (CGRS D) from the Natural Sciences and Engineering Research Council of Canada (NSERC). We acknowledge support from the SPHEREx project under a contract from the NASA/GODDARD Space Flight Center to the California Institute of Technology.

\bibliography{references}

@ARTICLE{2007A&A...464..399H,
       author = {{Hartlap}, J. and {Simon}, P. and {Schneider}, P.},
        title = "{Why your model parameter confidences might be too optimistic. Unbiased estimation of the inverse covariance matrix}",
      journal = {\aap},
         year = 2007,
        month = mar,
       volume = {464},
       number = {1},
        pages = {399-404},
          doi = {10.1051/0004-6361:20066170},
archivePrefix = {arXiv},
       eprint = {astro-ph/0608064},
 primaryClass = {astro-ph},
       adsurl = {https://ui.adsabs.harvard.edu/abs/2007A&A...464..399H}
}

@ARTICLE{Dalal+:2025,
       author = {{Dalal}, Neal and {Percival}, Will J.},
        title = "{Estimating non-gaussian bias using counts of tracers}",
      journal = {arXiv e-prints},
         year = 2025,
        month = mar,
          eid = {arXiv:2503.21024},
        pages = {arXiv:2503.21024},
          doi = {10.48550/arXiv.2503.21024},
archivePrefix = {arXiv},
       eprint = {2503.21024},
 primaryClass = {astro-ph.CO},
       adsurl = {https://ui.adsabs.harvard.edu/abs/2025arXiv250321024D}
}

@ARTICLE{Sullivan+:2025,
        author = {{Sullivan}, James M. and {Seljak}, Uro{\v{s}}},
        title = "{Local primordial non-Gaussian bias from time evolution}",
      journal = {\prd},
         year = 2025,
        month = oct,
       volume = {112},
       number = {8},
          eid = {083522},
        pages = {083522},
          doi = {10.1103/8jvj-66zc},
archivePrefix = {arXiv},
       eprint = {2503.21736},
 primaryClass = {astro-ph.CO},
       adsurl = {https://ui.adsabs.harvard.edu/abs/2025PhRvD.112h3522S}
}

@ARTICLE{Shiveshwarkar:2025+,
       author = {{Shiveshwarkar}, Charuhas and {Loverde}, Marilena and {Hirata}, Christopher M. and {Jamieson}, Drew},
        title = "{Where does non-Universality in Assembly Bias come from?}",
      journal = {arXiv e-prints},
         year = 2025,
        month = aug,
          eid = {arXiv:2508.11798},
        pages = {arXiv:2508.11798},
          doi = {10.48550/arXiv.2508.11798},
archivePrefix = {arXiv},
       eprint = {2508.11798},
 primaryClass = {astro-ph.CO},
       adsurl = {https://ui.adsabs.harvard.edu/abs/2025arXiv250811798S}
}

@ARTICLE{Hadzhiyska+:2025,
       author = {{Hadzhiyska}, Boryana and {Ferraro}, Simone},
        title = "{Refining localtype primordial non-Gaussianity: Sharpened b{\ensuremath{\phi}} constraints through bias expansion}",
      journal = {\prd},
         year = 2025,
        month = may,
       volume = {111},
       number = {10},
          eid = {103521},
        pages = {103521},
          doi = {10.1103/PhysRevD.111.103521},
archivePrefix = {arXiv},
       eprint = {2501.14873},
 primaryClass = {astro-ph.CO},
       adsurl = {https://ui.adsabs.harvard.edu/abs/2025PhRvD.111j3521H}
}

@ARTICLE{Perez+:2026,
       author = {{Perez}, Lucia A. and {Genel}, Shy and {Krause}, Elisabeth and {Somerville}, Rachel S.},
        title = "{The Impact of Galaxy Formation on Galaxy Biasing, and Implications for Primordial non-Gaussianity Constraints}",
      journal = {arXiv e-prints},
         year = 2026,
        month = feb,
          eid = {arXiv:2602.04987},
        pages = {arXiv:2602.04987},
          doi = {10.48550/arXiv.2602.04987},
archivePrefix = {arXiv},
       eprint = {2602.04987},
 primaryClass = {astro-ph.CO},
       adsurl = {https://ui.adsabs.harvard.edu/abs/2026arXiv260204987P}
}

@ARTICLE{Moore+:2026,
       author = {{Moore}, Anne and {Perez}, Lucia A. and {Krause}, Elisabeth},
        title = "{Informative Priors on Primordial Non-Gaussianity Bias $b_ϕ$ From Galaxy Formation}",
      journal = {arXiv e-prints},
         year = 2026,
        month = apr,
          eid = {arXiv:2604.21790},
        pages = {arXiv:2604.21790},
          doi = {10.48550/arXiv.2604.21790},
archivePrefix = {arXiv},
       eprint = {2604.21790},
 primaryClass = {astro-ph.CO},
       adsurl = {https://ui.adsabs.harvard.edu/abs/2026arXiv260421790M}
}

@ARTICLE{Bock+:2026ApJ...999..139B,
       author = {{Bock}, James J. and {Aboobaker}, Asad M. and {Adamo}, Joseph and {Akeson}, Rachel and {Alred}, John M. and {Alibay}, Farah and {Ashby}, Matthew L.~N. and {Bach}, Yoonsoo P. and {Bleem}, Lindsey E. and {Bolton}, Douglas and {Braun}, David F. and {Bruton}, Sean and {Bryan}, Sean A. and {Chang}, Tzu-Ching and {Chen}, Shuang-Shuang and {Cheng}, Yun-Ting and {Cheshire}, IV, James R. and {Chiang}, Yi-Kuan and {de Janvry}, Jean Choppin and {Condon}, Samuel and {Cook}, Walter R. and {Cooray}, Asantha and {Crill}, Brendan P. and {Cukierman}, Ari J. and {Dor{\'e}}, Olivier and {SPHEREx Collaboration}},
        title = "{The SPHEREx Satellite Mission}",
      journal = {\apj},
         year = 2026,
        month = mar,
       volume = {999},
       number = {1},
          eid = {139},
        pages = {139},
          doi = {10.3847/1538-4357/ae2be2},
archivePrefix = {arXiv},
       eprint = {2511.02985},
 primaryClass = {astro-ph.IM},
       adsurl = {https://ui.adsabs.harvard.edu/abs/2026ApJ...999..139B}
}

@ARTICLE{Wen+:2025PhRvD.112,
  title = {Separating angular and radial modes with spherical-Fourier Bessel power spectrum on all scales and implications for systematics mitigation},
  author = {Wen, Robin Y. and Gebhardt, Henry S. Grasshorn and Heinrich, Chen and Dor\'e, Olivier},
  journal = {Phys. Rev. D},
  volume = {112},
  issue = {6},
  pages = {063518},
  numpages = {22},
  year = {2025},
  month = {Sep},
  publisher = {American Physical Society},
  doi = {10.1103/hqc3-tbxc},
  url = {https://link.aps.org/doi/10.1103/hqc3-tbxc},
archivePrefix = {arXiv},
       eprint = {2506.06902},
 primaryClass = {astro-ph.CO},
       adsurl = {https://ui.adsabs.harvard.edu/abs/2025arXiv250606902W}
}

@ARTICLE{Tellarini+:2015,
       author = {{Tellarini}, Matteo and {Ross}, Ashley J. and {Tasinato}, Gianmassimo and {Wands}, David},
        title = "{Non-local bias in the halo bispectrum with primordial non-Gaussianity}",
      journal = {\jcap},
         year = 2015,
        month = jul,
       volume = {2015},
       number = {7},
        pages = {004-004},
          doi = {10.1088/1475-7516/2015/07/004},
archivePrefix = {arXiv},
       eprint = {1504.00324},
 primaryClass = {astro-ph.CO},
       adsurl = {https://ui.adsabs.harvard.edu/abs/2015JCAP...07..004T}
}

@ARTICLE{Wen+:2024PhRvD.110l3501W,
       author = {{Wen}, Robin Y. and {Gebhardt}, Henry S. Grasshorn and {Heinrich}, Chen and {Dor{\'e}}, Olivier},
        title = "{Linear relativistic corrections in the spherical Fourier-Bessel power spectrum}",
      journal = {\prd},
         year = 2024,
        month = dec,
       volume = {110},
       number = {12},
          eid = {123501},
        pages = {123501},
          doi = {10.1103/PhysRevD.110.123501},
archivePrefix = {arXiv},
       eprint = {2407.02753},
 primaryClass = {astro-ph.CO},
       adsurl = {https://ui.adsabs.harvard.edu/abs/2024PhRvD.110l3501W}
}

@ARTICLE{Khek+:2024PhRvD.110f3524K,
       author = {{Khek}, Brandon and {Gebhardt}, Henry Grasshorn and {Dor{\'e}}, Olivier},
        title = "{Fast theoretical predictions for spherical Fourier analysis of large-scale structures}",
      journal = {\prd},
         year = 2024,
        month = sep,
       volume = {110},
       number = {6},
          eid = {063524},
        pages = {063524},
          doi = {10.1103/PhysRevD.110.063524},
archivePrefix = {arXiv},
       eprint = {2212.05760},
 primaryClass = {astro-ph.CO},
       adsurl = {https://ui.adsabs.harvard.edu/abs/2024PhRvD.110f3524K}
}

@ARTICLE{GrasshornGebhardt+:2024PhRvD.109h3502G,
       author = {{Grasshorn Gebhardt}, Henry S. and {Dor{\'e}}, Olivier},
        title = "{Validation of spherical Fourier-Bessel power spectrum analysis with log-normal simulations and eBOSS DR16 LRG EZmocks}",
      journal = {\prd},
         year = 2024,
        month = apr,
       volume = {109},
       number = {8},
          eid = {083502},
        pages = {083502},
          doi = {10.1103/PhysRevD.109.083502},
archivePrefix = {arXiv},
       eprint = {2310.17677},
 primaryClass = {astro-ph.IM},
       adsurl = {https://ui.adsabs.harvard.edu/abs/2024PhRvD.109h3502G}
}

@ARTICLE{GrasshornGebhardt+:2021PhRvD.104l3548G,
       author = {{Grasshorn Gebhardt}, Henry S. and {Dor{\'e}}, Olivier},
        title = "{Fabulous code for spherical Fourier-Bessel decomposition}",
      journal = {\prd},
         year = 2021,
        month = dec,
       volume = {104},
       number = {12},
          eid = {123548},
        pages = {123548},
          doi = {10.1103/PhysRevD.104.123548},
archivePrefix = {arXiv},
       eprint = {2102.10079},
 primaryClass = {astro-ph.CO},
       adsurl = {https://ui.adsabs.harvard.edu/abs/2021PhRvD.104l3548G}
}

@ARTICLE{Eifler+:2021MNRAS.507.1746E,
       author = {{Eifler}, Tim and {Miyatake}, Hironao and {Krause}, Elisabeth and {Heinrich}, Chen and {Miranda}, Vivian and {Hirata}, Christopher and {Xu}, Jiachuan and {Hemmati}, Shoubaneh and {Simet}, Melanie and {Capak}, Peter and {Choi}, Ami and {Dor{\'e}}, Olivier and {Doux}, Cyrille and {Fang}, Xiao and {Hounsell}, Rebekah and {Huff}, Eric and {Huang}, Hung-Jin and {Jarvis}, Mike and {Kruk}, Jeffrey and {Masters}, Dan and {Rozo}, Eduardo and {Scolnic}, Dan and {Spergel}, David N. and {Troxel}, Michael and {von der Linden}, Anja and {Wang}, Yun and {Weinberg}, David H. and {Wenzl}, Lukas and {Wu}, Hao-Yi},
        title = "{Cosmology with the Roman Space Telescope - multiprobe strategies}",
      journal = {\mnras},
         year = 2021,
        month = oct,
       volume = {507},
       number = {2},
        pages = {1746-1761},
          doi = {10.1093/mnras/stab1762},
archivePrefix = {arXiv},
       eprint = {2004.05271},
 primaryClass = {astro-ph.CO},
       adsurl = {https://ui.adsabs.harvard.edu/abs/2021MNRAS.507.1746E}
}

@software{Tomasi+:2021ascl.soft09028T,
       author = {{Tomasi}, Maurizio and {Li}, Zack},
        title = "{Healpix.jl: Julia-only port of the HEALPix library}",
 howpublished = {Astrophysics Source Code Library, record ascl:2109.028},
         year = 2021,
        month = sep,
          eid = {ascl:2109.028},
       adsurl = {https://ui.adsabs.harvard.edu/abs/2021ascl.soft09028T}
}

@ARTICLE{Weaverdyck+:2021MNRAS.503.5061W,
       author = {{Weaverdyck}, Noah and {Huterer}, Dragan},
        title = "{Mitigating contamination in LSS surveys: a comparison of methods}",
      journal = {\mnras},
         year = 2021,
        month = may,
       volume = {503},
       number = {4},
        pages = {5061-5084},
          doi = {10.1093/mnras/stab709},
archivePrefix = {arXiv},
       eprint = {2007.14499},
 primaryClass = {astro-ph.CO},
       adsurl = {https://ui.adsabs.harvard.edu/abs/2021MNRAS.503.5061W}
}

@ARTICLE{Zhao+:2021MNRAS.503.1149Z,
       author = {{Zhao}, Cheng and {Chuang}, Chia-Hsun and {Bautista}, Julian and {de Mattia}, Arnaud and {Raichoor}, Anand and {Ross}, Ashley J. and {Hou}, Jiamin and {Neveux}, Richard and {Tao}, Charling and {Burtin}, Etienne and {Dawson}, Kyle S. and {de la Torre}, Sylvain and {Gil-Mar{\'\i}n}, H{\'e}ctor and {Kneib}, Jean-Paul and {Percival}, Will J. and {Rossi}, Graziano and {Tamone}, Am{\'e}lie and {Tinker}, Jeremy L. and {Zhao}, Gong-Bo and {Alam}, Shadab and {Mueller}, Eva-Maria},
        title = "{The completed SDSS-IV extended Baryon Oscillation Spectroscopic Survey: 1000 multi-tracer mock catalogues with redshift evolution and systematics for galaxies and quasars of the final data release}",
      journal = {\mnras},
         year = 2021,
        month = may,
       volume = {503},
       number = {1},
        pages = {1149-1173},
          doi = {10.1093/mnras/stab510},
archivePrefix = {arXiv},
       eprint = {2007.08997},
 primaryClass = {astro-ph.CO},
       adsurl = {https://ui.adsabs.harvard.edu/abs/2021MNRAS.503.1149Z}
}

@ARTICLE{Ross+:2020MNRAS.498.2354R,
       author = {{Ross}, Ashley J. and {Bautista}, Julian and {Tojeiro}, Rita and {Alam}, Shadab and {Bailey}, Stephen and {Burtin}, Etienne and {Comparat}, Johan and {Dawson}, Kyle S. and {de Mattia}, Arnaud and {du Mas des Bourboux}, H{\'e}lion and {Gil-Mar{\'\i}n}, H{\'e}ctor and {Hou}, Jiamin and {Kong}, Hui and {Lyke}, Brad W. and {Mohammad}, Faizan G. and {Moustakas}, John and {Mueller}, Eva-Maria and {Myers}, Adam D. and {Percival}, Will J. and {Raichoor}, Anand and {Rezaie}, Mehdi and {Seo}, Hee-Jong and {Smith}, Alex and {Tinker}, Jeremy L. and {Zarrouk}, Pauline and {Zhao}, Cheng and {Zhao}, Gong-Bo and {Bizyaev}, Dmitry and {Brinkmann}, Jonathan and {Brownstein}, Joel R. and {Rosell}, Aurelio Carnero and {Chabanier}, Sol{\`e}ne and {Choi}, Peter D. and {Chuang}, Chia-Hsun and {Cruz-Gonzalez}, Irene and {de la Macorra}, Axel and {de la Torre}, Sylvain and {Escoffier}, Stephanie and {Fromenteau}, Sebastien and {Higley}, Alexandra and {Jullo}, Eric and {Kneib}, Jean-Paul and {McLane}, Jacob N. and {Mu{\~n}oz-Guti{\'e}rrez}, Andrea and {Neveux}, Richard and {Newman}, Jeffrey A. and {Nitschelm}, Christian and {Palanque-Delabrouille}, Nathalie and {Paviot}, Romain and {Pullen}, Anthony R. and {Rossi}, Graziano and {Ruhlmann-Kleider}, Vanina and {Schneider}, Donald P. and {Maga{\~n}a}, Mariana Vargas and {Vivek}, M. and {Zhang}, Yucheng},
        title = "{The Completed SDSS-IV extended Baryon Oscillation Spectroscopic Survey: Large-scale structure catalogues for cosmological analysis}",
      journal = {\mnras},
         year = 2020,
        month = oct,
       volume = {498},
       number = {2},
        pages = {2354-2371},
          doi = {10.1093/mnras/staa2416},
archivePrefix = {arXiv},
       eprint = {2007.09000},
 primaryClass = {astro-ph.CO},
       adsurl = {https://ui.adsabs.harvard.edu/abs/2020MNRAS.498.2354R}
}

@ARTICLE{Wang+:2020JCAP...10..022W,
       author = {{Wang}, M.~S. and {Avila}, S. and {Bianchi}, D. and {Crittenden}, R. and {Percival}, W.~J.},
        title = "{Hybrid-basis inference for large-scale galaxy clustering: combining spherical and Cartesian Fourier analyses}",
      journal = {\jcap},
         year = 2020,
        month = oct,
       volume = {2020},
       number = {10},
          eid = {022},
        pages = {022},
          doi = {10.1088/1475-7516/2020/10/022},
archivePrefix = {arXiv},
       eprint = {2007.14962},
 primaryClass = {astro-ph.CO},
       adsurl = {https://ui.adsabs.harvard.edu/abs/2020JCAP...10..022W}
}

@ARTICLE{PlanckCollaboration:2020A&A...641A...6P,
       author = {{Planck Collaboration} and {Aghanim}, N. and {Akrami}, Y. and {Ashdown}, M. and {Aumont}, J. and {Baccigalupi}, C. and {Ballardini}, M. and {Banday}, A.~J. and {Barreiro}, R.~B. and {Bartolo}, N. and {Basak}, S. and {Battye}, R. and {Benabed}, K. and {Bernard}, J. -P. and {Bersanelli}, M. and {Bielewicz}, P. and {Bock}, J.~J. and {Bond}, J.~R. and {Borrill}, J. and {Bouchet}, F.~R. and {Boulanger}, F. and {Bucher}, M. and {Burigana}, C. and {Butler}, R.~C. and {Calabrese}, E. and {Cardoso}, J. -F. and {Carron}, J. and {Challinor}, A. and {Chiang}, H.~C. and {Chluba}, J. and {Colombo}, L.~P.~L. and {Combet}, C. and {Contreras}, D. and {Crill}, B.~P. and {Cuttaia}, F. and {de Bernardis}, P. and {de Zotti}, G. and {Delabrouille}, J. and {Delouis}, J. -M. and {Di Valentino}, E. and {Diego}, J.~M. and {Dor{\'e}}, O. and {Douspis}, M. and {Ducout}, A. and {Dupac}, X. and {Dusini}, S. and {Efstathiou}, G. and {Elsner}, F. and {En{\ss}lin}, T.~A. and {Eriksen}, H.~K. and {Fantaye}, Y. and {Farhang}, M. and {Fergusson}, J. and {Fernandez-Cobos}, R. and {Finelli}, F. and {Forastieri}, F. and {Frailis}, M. and {Fraisse}, A.~A. and {Franceschi}, E. and {Frolov}, A. and {Galeotta}, S. and {Galli}, S. and {Ganga}, K. and {G{\'e}nova-Santos}, R.~T. and {Gerbino}, M. and {Ghosh}, T. and {Gonz{\'a}lez-Nuevo}, J. and {G{\'o}rski}, K.~M. and {Gratton}, S. and {Gruppuso}, A. and {Gudmundsson}, J.~E. and {Hamann}, J. and {Handley}, W. and {Hansen}, F.~K. and {Herranz}, D. and {Hildebrandt}, S.~R. and {Hivon}, E. and {Huang}, Z. and {Jaffe}, A.~H. and {Jones}, W.~C. and {Karakci}, A. and {Keih{\"a}nen}, E. and {Keskitalo}, R. and {Kiiveri}, K. and {Kim}, J. and {Kisner}, T.~S. and {Knox}, L. and {Krachmalnicoff}, N. and {Kunz}, M. and {Kurki-Suonio}, H. and {Lagache}, G. and {Lamarre}, J. -M. and {Lasenby}, A. and {Lattanzi}, M. and {Lawrence}, C.~R. and {Le Jeune}, M. and {Lemos}, P. and {Lesgourgues}, J. and {Levrier}, F. and {Lewis}, A. and {Liguori}, M. and {Lilje}, P.~B. and {Lilley}, M. and {Lindholm}, V. and {L{\'o}pez-Caniego}, M. and {Lubin}, P.~M. and {Ma}, Y. -Z. and {Mac{\'\i}as-P{\'e}rez}, J.~F. and {Maggio}, G. and {Maino}, D. and {Mandolesi}, N. and {Mangilli}, A. and {Marcos-Caballero}, A. and {Maris}, M. and {Martin}, P.~G. and {Martinelli}, M. and {Mart{\'\i}nez-Gonz{\'a}lez}, E. and {Matarrese}, S. and {Mauri}, N. and {McEwen}, J.~D. and {Meinhold}, P.~R. and {Melchiorri}, A. and {Mennella}, A. and {Migliaccio}, M. and {Millea}, M. and {Mitra}, S. and {Miville-Desch{\^e}nes}, M. -A. and {Molinari}, D. and {Montier}, L. and {Morgante}, G. and {Moss}, A. and {Natoli}, P. and {N{\o}rgaard-Nielsen}, H.~U. and {Pagano}, L. and {Paoletti}, D. and {Partridge}, B. and {Patanchon}, G. and {Peiris}, H.~V. and {Perrotta}, F. and {Pettorino}, V. and {Piacentini}, F. and {Polastri}, L. and {Polenta}, G. and {Puget}, J. -L. and {Rachen}, J.~P. and {Reinecke}, M. and {Remazeilles}, M. and {Renzi}, A. and {Rocha}, G. and {Rosset}, C. and {Roudier}, G. and {Rubi{\~n}o-Mart{\'\i}n}, J.~A. and {Ruiz-Granados}, B. and {Salvati}, L. and {Sandri}, M. and {Savelainen}, M. and {Scott}, D. and {Shellard}, E.~P.~S. and {Sirignano}, C. and {Sirri}, G. and {Spencer}, L.~D. and {Sunyaev}, R. and {Suur-Uski}, A. -S. and {Tauber}, J.~A. and {Tavagnacco}, D. and {Tenti}, M. and {Toffolatti}, L. and {Tomasi}, M. and {Trombetti}, T. and {Valenziano}, L. and {Valiviita}, J. and {Van Tent}, B. and {Vibert}, L. and {Vielva}, P. and {Villa}, F. and {Vittorio}, N. and {Wandelt}, B.~D. and {Wehus}, I.~K. and {White}, M. and {White}, S.~D.~M. and {Zacchei}, A. and {Zonca}, A.},
        title = "{Planck 2018 results. VI. Cosmological parameters}",
      journal = {\aap},
         year = 2020,
        month = sep,
       volume = {641},
          eid = {A6},
        pages = {A6},
          doi = {10.1051/0004-6361/201833910},
archivePrefix = {arXiv},
       eprint = {1807.06209},
 primaryClass = {astro-ph.CO},
       adsurl = {https://ui.adsabs.harvard.edu/abs/2020A&A...641A...6P}
}

@ARTICLE{Castorina+:2019JCAP...09..010C,
       author = {{Castorina}, Emanuele and {Hand}, Nick and {Seljak}, Uro{\v{s}} and {Beutler}, Florian and {Chuang}, Chia-Hsun and {Zhao}, Cheng and {Gil-Mar{\'\i}n}, H{\'e}ctor and {Percival}, Will J. and {Ross}, Ashley J. and {Choi}, Peter Doohyun and {Dawson}, Kyle and {de la Macorra}, Axel and {Rossi}, Graziano and {Ruggeri}, Rossana and {Schneider}, Donald and {Zhao}, Gong-Bo},
        title = "{Redshift-weighted constraints on primordial non-Gaussianity from the clustering of the eBOSS DR14 quasars in Fourier space}",
      journal = {\jcap},
         year = 2019,
        month = sep,
       volume = {2019},
       number = {9},
          eid = {010},
        pages = {010},
          doi = {10.1088/1475-7516/2019/09/010},
archivePrefix = {arXiv},
       eprint = {1904.08859},
 primaryClass = {astro-ph.CO},
       adsurl = {https://ui.adsabs.harvard.edu/abs/2019JCAP...09..010C}
}

@ARTICLE{Qin+:2019MNRAS.487.5235Q,
       author = {{Qin}, Fei and {Howlett}, Cullan and {Staveley-Smith}, Lister},
        title = "{The redshift-space momentum power spectrum - II. Measuring the growth rate from the combined 2MTF and 6dFGSv surveys}",
      journal = {\mnras},
         year = 2019,
        month = aug,
       volume = {487},
       number = {4},
        pages = {5235-5247},
          doi = {10.1093/mnras/stz1576},
archivePrefix = {arXiv},
       eprint = {1906.02874},
 primaryClass = {astro-ph.CO},
       adsurl = {https://ui.adsabs.harvard.edu/abs/2019MNRAS.487.5235Q}
}

@ARTICLE{deMattia+:2019JCAP...08..036D,
       author = {{de Mattia}, Arnaud and {Ruhlmann-Kleider}, Vanina},
        title = "{Integral constraints in spectroscopic surveys}",
      journal = {\jcap},
         year = 2019,
        month = aug,
       volume = {2019},
       number = {8},
          eid = {036},
        pages = {036},
          doi = {10.1088/1475-7516/2019/08/036},
archivePrefix = {arXiv},
       eprint = {1904.08851},
 primaryClass = {astro-ph.CO},
       adsurl = {https://ui.adsabs.harvard.edu/abs/2019JCAP...08..036D}
}

@ARTICLE{Samushia:2019,
       author = {{Samushia}, Lado},
        title = "{Proper Fourier decomposition formalism for cosmological fields in spherical shells}",
      journal = {arXiv e-prints},
         year = 2019,
        month = jun,
          eid = {arXiv:1906.05866},
        pages = {arXiv:1906.05866},
          doi = {10.48550/arXiv.1906.05866},
archivePrefix = {arXiv},
       eprint = {1906.05866},
 primaryClass = {astro-ph.CO},
       adsurl = {https://ui.adsabs.harvard.edu/abs/2019arXiv190605866S}
}

@ARTICLE{Dore+:2014arXiv1412.4872D,
       author = {{Dor{\'e}}, Olivier and {Bock}, Jamie and {Ashby}, Matthew and {Capak}, Peter and {Cooray}, Asantha and {de Putter}, Roland and {Eifler}, Tim and {Flagey}, Nicolas and {Gong}, Yan and {Habib}, Salman and {Heitmann}, Katrin and {Hirata}, Chris and {Jeong}, Woong-Seob and {Katti}, Raj and {Korngut}, Phil and {Krause}, Elisabeth and {Lee}, Dae-Hee and {Masters}, Daniel and {Mauskopf}, Phil and {Melnick}, Gary and {Mennesson}, Bertrand and {Nguyen}, Hien and {{\"O}berg}, Karin and {Pullen}, Anthony and {Raccanelli}, Alvise and {Smith}, Roger and {Song}, Yong-Seon and {Tolls}, Volker and {Unwin}, Steve and {Venumadhav}, Tejaswi and {Viero}, Marco and {Werner}, Mike and {Zemcov}, Mike},
        title = "{Cosmology with the SPHEREX All-Sky Spectral Survey}",
      journal = {arXiv e-prints},
         year = 2014,
        month = dec,
          eid = {arXiv:1412.4872},
        pages = {arXiv:1412.4872},
          doi = {10.48550/arXiv.1412.4872},
archivePrefix = {arXiv},
       eprint = {1412.4872},
 primaryClass = {astro-ph.CO},
       adsurl = {https://ui.adsabs.harvard.edu/abs/2014arXiv1412.4872D}
}

@ARTICLE{Leistedt+:2014PhRvL.113v1301L,
       author = {{Leistedt}, Boris and {Peiris}, Hiranya V. and {Roth}, Nina},
        title = "{Constraints on Primordial Non-Gaussianity from 800 000 Photometric Quasars}",
      journal = {\prl},
         year = 2014,
        month = nov,
       volume = {113},
       number = {22},
          eid = {221301},
        pages = {221301},
          doi = {10.1103/PhysRevLett.113.221301},
archivePrefix = {arXiv},
       eprint = {1405.4315},
 primaryClass = {astro-ph.CO},
       adsurl = {https://ui.adsabs.harvard.edu/abs/2014PhRvL.113v1301L}
}

@ARTICLE{Leistedt+:2012A&A...540A..60L,
       author = {{Leistedt}, B. and {Rassat}, A. and {R{\'e}fr{\'e}gier}, A. and {Starck}, J. -L.},
        title = "{3DEX: a code for fast spherical Fourier-Bessel decomposition of 3D surveys}",
      journal = {\aap},
         year = 2012,
        month = apr,
       volume = {540},
          eid = {A60},
        pages = {A60},
          doi = {10.1051/0004-6361/201118463},
archivePrefix = {arXiv},
       eprint = {1111.3591},
 primaryClass = {astro-ph.CO},
       adsurl = {https://ui.adsabs.harvard.edu/abs/2012A&A...540A..60L}
}

@software{Leistedt+:2011ascl.soft11011L,
       author = {{Leistedt}, B. and {Rassat}, A. and {Refregier}, A. and {Starck}, J. -L.},
        title = "{3DEX: Fast Fourier-Bessel Decomposition of Spherical 3D Surveys}",
 howpublished = {Astrophysics Source Code Library, record ascl:1111.011},
         year = 2011,
        month = nov,
          eid = {ascl:1111.011},
       adsurl = {https://ui.adsabs.harvard.edu/abs/2011ascl.soft11011L}
}

@ARTICLE{Bonvin+:2011PhRvD..84f3505B,
       author = {{Bonvin}, Camille and {Durrer}, Ruth},
        title = "{What galaxy surveys really measure}",
      journal = {\prd},
         year = 2011,
        month = sep,
       volume = {84},
       number = {6},
          eid = {063505},
        pages = {063505},
          doi = {10.1103/PhysRevD.84.063505},
archivePrefix = {arXiv},
       eprint = {1105.5280},
 primaryClass = {astro-ph.CO},
       adsurl = {https://ui.adsabs.harvard.edu/abs/2011PhRvD..84f3505B}
}

@ARTICLE{Challinor+:2011PhRvD..84d3516C,
       author = {{Challinor}, Anthony and {Lewis}, Antony},
        title = "{Linear power spectrum of observed source number counts}",
      journal = {\prd},
         year = 2011,
        month = aug,
       volume = {84},
       number = {4},
          eid = {043516},
        pages = {043516},
          doi = {10.1103/PhysRevD.84.043516},
archivePrefix = {arXiv},
       eprint = {1105.5292},
 primaryClass = {astro-ph.CO},
       adsurl = {https://ui.adsabs.harvard.edu/abs/2011PhRvD..84d3516C}
}

@ARTICLE{Slosar+:2008JCAP...08..031S,
       author = {{Slosar}, An{\v{z}}e and {Hirata}, Christopher and {Seljak}, Uro{\v{s}} and {Ho}, Shirley and {Padmanabhan}, Nikhil},
        title = "{Constraints on local primordial non-Gaussianity from large scale structure}",
      journal = {\jcap},
         year = 2008,
        month = aug,
       volume = {2008},
       number = {8},
          eid = {031},
        pages = {031},
          doi = {10.1088/1475-7516/2008/08/031},
archivePrefix = {arXiv},
       eprint = {0805.3580},
 primaryClass = {astro-ph},
       adsurl = {https://ui.adsabs.harvard.edu/abs/2008JCAP...08..031S}
}

@ARTICLE{Dalal+:2008PhRvD..77l3514D,
       author = {{Dalal}, Neal and {Dor{\'e}}, Olivier and {Huterer}, Dragan and {Shirokov}, Alexander},
        title = "{Imprints of primordial non-Gaussianities on large-scale structure: Scale-dependent bias and abundance of virialized objects}",
      journal = {\prd},
         year = 2008,
        month = jun,
       volume = {77},
       number = {12},
          eid = {123514},
        pages = {123514},
          doi = {10.1103/PhysRevD.77.123514},
archivePrefix = {arXiv},
       eprint = {0710.4560},
 primaryClass = {astro-ph},
       adsurl = {https://ui.adsabs.harvard.edu/abs/2008PhRvD..77l3514D}
}

@ARTICLE{Yamamoto+:2006PASJ...58...93Y,
       author = {{Yamamoto}, Kazuhiro and {Nakamichi}, Masashi and {Kamino}, Akinari and {Bassett}, Bruce A. and {Nishioka}, Hiroaki},
        title = "{A Measurement of the Quadrupole Power Spectrum in the Clustering of the 2dF QSO Survey}",
      journal = {\pasj},
         year = 2006,
        month = feb,
       volume = {58},
        pages = {93-102},
          doi = {10.1093/pasj/58.1.93},
archivePrefix = {arXiv},
       eprint = {astro-ph/0505115},
 primaryClass = {astro-ph},
       adsurl = {https://ui.adsabs.harvard.edu/abs/2006PASJ...58...93Y}
}

@ARTICLE{Gorski+:2005ApJ...622..759G,
       author = {{G{\'o}rski}, K.~M. and {Hivon}, E. and {Banday}, A.~J. and {Wandelt}, B.~D. and {Hansen}, F.~K. and {Reinecke}, M. and {Bartelmann}, M.},
        title = "{HEALPix: A Framework for High-Resolution Discretization and Fast Analysis of Data Distributed on the Sphere}",
      journal = {\apj},
         year = 2005,
        month = apr,
       volume = {622},
       number = {2},
        pages = {759-771},
          doi = {10.1086/427976},
archivePrefix = {arXiv},
       eprint = {astro-ph/0409513},
 primaryClass = {astro-ph},
       adsurl = {https://ui.adsabs.harvard.edu/abs/2005ApJ...622..759G}
}

@INPROCEEDINGS{Komatsu+:2002nmgm.meet.2009K,
       author = {{Komatsu}, Eiichiro and {Spergel}, David N.},
        title = "{The Cosmic Microwave Background Bispectrum as a Test of the Physics of Inflation and Probe of the Astrophysics of the Low-Redshift Universe}",
    booktitle = {The Ninth Marcel Grossmann Meeting},
         year = 2002,
       editor = {{Gurzadyan}, Vahe G. and {Jantzen}, Robert T. and {Ruffini}, Remo},
        month = dec,
        pages = {2009-2010},
          doi = {10.1142/9789812777386_0459},
archivePrefix = {arXiv},
       eprint = {astro-ph/0012197},
 primaryClass = {astro-ph},
       adsurl = {https://ui.adsabs.harvard.edu/abs/2002nmgm.meet.2009K}
}

@ARTICLE{Feldman+:1994ApJ...426...23F,
       author = {{Feldman}, Hume A. and {Kaiser}, Nick and {Peacock}, John A.},
        title = "{Power-Spectrum Analysis of Three-dimensional Redshift Surveys}",
      journal = {\apj},
         year = 1994,
        month = may,
       volume = {426},
        pages = {23},
          doi = {10.1086/174036},
archivePrefix = {arXiv},
       eprint = {astro-ph/9304022},
 primaryClass = {astro-ph},
       adsurl = {https://ui.adsabs.harvard.edu/abs/1994ApJ...426...23F}
}

@ARTICLE{Wen+:2026,
       author = {{Wen}, Robin Y. and {Grasshorn Gebhardt}, Henry S. and {Heinrich}, Chen and {Dor{\'e}}, Olivier},
        title = "{Large-scale Modeling of the Observed Power Spectrum Multipoles}",
      journal = {arXiv e-prints},
         year = 2026,
        month = jan,
          eid = {arXiv:2601.19438},
        pages = {arXiv:2601.19438},
archivePrefix = {arXiv},
       eprint = {2601.19438},
 primaryClass = {astro-ph.CO},
       adsurl = {https://ui.adsabs.harvard.edu/abs/2026arXiv260119438W}
}

@article{Roberts+:doi:10.1198/jcgs.2009.06134,
author = {Gareth O. Roberts and Jeffrey S. Rosenthal},
title = {Examples of Adaptive MCMC},
journal = {Journal of Computational and Graphical Statistics},
volume = {18},
number = {2},
pages = {349-367},
year  = {2009},
publisher = {Taylor & Francis},
doi = {10.1198/jcgs.2009.06134},

URL = { 
    
        https://doi.org/10.1198/jcgs.2009.06134
    
    

},
eprint = { 
    
        https://doi.org/10.1198/jcgs.2009.06134
    
    

}
}

@ARTICLE{Ivezic+:2019ApJ...873..111I,
       author = {{Ivezi{\'c}}, {\v{Z}}eljko and {Kahn}, Steven M. and {Tyson}, J. Anthony and {Abel}, Bob and {Acosta}, Emily and {Allsman}, Robyn and {Alonso}, David and {AlSayyad}, Yusra and {Anderson}, Scott F. and {Andrew}, John and {Angel}, James Roger P. and {Angeli}, George Z. and {Ansari}, Reza and {Antilogus}, Pierre and {Araujo}, Constanza and {Armstrong}, Robert and {Arndt}, Kirk T. and {Astier}, Pierre and {Aubourg}, {\'E}ric and {Auza}, Nicole and {Axelrod}, Tim S. and {Bard}, Deborah J. and {Barr}, Jeff D. and {Barrau}, Aurelian and {Bartlett}, James G. and {Bauer}, Amanda E. and {Bauman}, Brian J. and {Baumont}, Sylvain and {Bechtol}, Ellen and {Bechtol}, Keith and {Becker}, Andrew C. and {Becla}, Jacek and {Beldica}, Cristina and {Bellavia}, Steve and {Bianco}, Federica B. and {Biswas}, Rahul and {Blanc}, Guillaume and {Blazek}, Jonathan and {Blandford}, Roger D. and {Bloom}, Josh S. and {Bogart}, Joanne and {Bond}, Tim W. and {Booth}, Michael T. and {Borgland}, Anders W. and {Borne}, Kirk and {Bosch}, James F. and {Boutigny}, Dominique and {Brackett}, Craig A. and {Bradshaw}, Andrew and {Brandt}, William Nielsen and {Brown}, Michael E. and {Bullock}, James S. and {Burchat}, Patricia and {Burke}, David L. and {Cagnoli}, Gianpietro and {Calabrese}, Daniel and {Callahan}, Shawn and {Callen}, Alice L. and {Carlin}, Jeffrey L. and {Carlson}, Erin L. and {Chandrasekharan}, Srinivasan and {Charles-Emerson}, Glenaver and {Chesley}, Steve and {Cheu}, Elliott C. and {Chiang}, Hsin-Fang and {Chiang}, James and {Chirino}, Carol and {Chow}, Derek and {Ciardi}, David R. and {Claver}, Charles F. and {Cohen-Tanugi}, Johann and {Cockrum}, Joseph J. and {Coles}, Rebecca and {Connolly}, Andrew J. and {Cook}, Kem H. and {Cooray}, Asantha and {Covey}, Kevin R. and {Cribbs}, Chris and {Cui}, Wei and {Cutri}, Roc and {Daly}, Philip N. and {Daniel}, Scott F. and {Daruich}, Felipe and {Daubard}, Guillaume and {Daues}, Greg and {Dawson}, William and {Delgado}, Francisco and {Dellapenna}, Alfred and {de Peyster}, Robert and {de Val-Borro}, Miguel and {Digel}, Seth W. and {Doherty}, Peter and {Dubois}, Richard and {Dubois-Felsmann}, Gregory P. and {Durech}, Josef and {Economou}, Frossie and {Eifler}, Tim and {Eracleous}, Michael and {Emmons}, Benjamin L. and {Fausti Neto}, Angelo and {Ferguson}, Henry and {Figueroa}, Enrique and {Fisher-Levine}, Merlin and {Focke}, Warren and {Foss}, Michael D. and {Frank}, James and {Freemon}, Michael D. and {Gangler}, Emmanuel and {Gawiser}, Eric and {Geary}, John C. and {Gee}, Perry and {Geha}, Marla and {Gessner}, Charles J.~B. and {Gibson}, Robert R. and {Gilmore}, D. Kirk and {Glanzman}, Thomas and {Glick}, William and {Goldina}, Tatiana and {Goldstein}, Daniel A. and {Goodenow}, Iain and {Graham}, Melissa L. and {Gressler}, William J. and {Gris}, Philippe and {Guy}, Leanne P. and {Guyonnet}, Augustin and {Haller}, Gunther and {Harris}, Ron and {Hascall}, Patrick A. and {Haupt}, Justine and {Hernandez}, Fabio and {Herrmann}, Sven and {Hileman}, Edward and {Hoblitt}, Joshua and {Hodgson}, John A. and {Hogan}, Craig and {Howard}, James D. and {Huang}, Dajun and {Huffer}, Michael E. and {Ingraham}, Patrick and {Innes}, Walter R. and {Jacoby}, Suzanne H. and {Jain}, Bhuvnesh and {Jammes}, Fabrice and {Jee}, M. James and {Jenness}, Tim and {Jernigan}, Garrett and {Jevremovi{\'c}}, Darko and {Johns}, Kenneth and {Johnson}, Anthony S. and {Johnson}, Margaret W.~G. and {Jones}, R. Lynne and {Juramy-Gilles}, Claire and {Juri{\'c}}, Mario and {Kalirai}, Jason S. and {Kallivayalil}, Nitya J. and {Kalmbach}, Bryce and {Kantor}, Jeffrey P. and {Karst}, Pierre and {Kasliwal}, Mansi M. and {Kelly}, Heather and {Kessler}, Richard and {Kinnison}, Veronica and {Kirkby}, David and {Knox}, Lloyd and {Kotov}, Ivan V. and {Krabbendam}, Victor L. and {Krughoff}, K. Simon and {Kub{\'a}nek}, Petr and {Kuczewski}, John and {Kulkarni}, Shri and {Ku}, John and {Kurita}, Nadine R. and {Lage}, Craig S. and {Lambert}, Ron and {Lange}, Travis and {Langton}, J. Brian and {Le Guillou}, Laurent and {Levine}, Deborah and {Liang}, Ming and {Lim}, Kian-Tat and {Lintott}, Chris J. and {Long}, Kevin E. and {Lopez}, Margaux and {Lotz}, Paul J. and {Lupton}, Robert H. and {Lust}, Nate B. and {MacArthur}, Lauren A. and {Mahabal}, Ashish and {Mandelbaum}, Rachel and {Markiewicz}, Thomas W. and {Marsh}, Darren S. and {Marshall}, Philip J. and {Marshall}, Stuart and {May}, Morgan and {McKercher}, Robert and {McQueen}, Michelle and {Meyers}, Joshua and {Migliore}, Myriam and {Miller}, Michelle and {Mills}, David J.},
        title = "{LSST: From Science Drivers to Reference Design and Anticipated Data Products}",
      journal = {\apj},
         year = 2019,
        month = mar,
       volume = {873},
       number = {2},
          eid = {111},
        pages = {111},
          doi = {10.3847/1538-4357/ab042c},
archivePrefix = {arXiv},
       eprint = {0805.2366},
 primaryClass = {astro-ph},
       adsurl = {https://ui.adsabs.harvard.edu/abs/2019ApJ...873..111I}
}

@ARTICLE{Modi+:2019JCAP...11..023M,
       author = {{Modi}, Chirag and {White}, Martin and {Slosar}, An{\v{z}}e and {Castorina}, Emanuele},
        title = "{Reconstructing large-scale structure with neutral hydrogen surveys}",
      journal = {\jcap},
         year = 2019,
        month = nov,
       volume = {2019},
       number = {11},
          eid = {023},
        pages = {023},
          doi = {10.1088/1475-7516/2019/11/023},
archivePrefix = {arXiv},
       eprint = {1907.02330},
 primaryClass = {astro-ph.CO},
       adsurl = {https://ui.adsabs.harvard.edu/abs/2019JCAP...11..023M}
}

@ARTICLE{EuclidCollaboration:2025A&A...697A...1E,
       author = {{Euclid Collaboration} and {Mellier}, Y. and {Abdurro'uf} and {Acevedo Barroso}, J.~A. and {Ach{\'u}carro}, A. and {Adamek}, J. and {Adam}, R. and {Addison}, G.~E. and {Aghanim}, N. and {Aguena}, M. and {Ajani}, V. and {Akrami}, Y. and {Al-Bahlawan}, A. and {Alavi}, A. and {Albuquerque}, I.~S. and {Alestas}, G. and {Alguero}, G. and {Allaoui}, A. and {Allen}, S.~W. and {Allevato}, V. and {Alonso-Tetilla}, A.~V. and {Altieri}, B. and {Alvarez-Candal}, A. and {Alvi}, S. and {Amara}, A. and {Amendola}, L. and {Amiaux}, J. and {Andika}, I.~T. and {Andreon}, S. and {Andrews}, A. and {Angora}, G. and {Angulo}, R.~E. and {Annibali}, F. and {Anselmi}, A. and {Anselmi}, S. and {Arcari}, S. and {Archidiacono}, M. and {Aric{\`o}}, G. and {Arnaud}, M. and {Arnouts}, S. and {Asgari}, M. and {Asorey}, J. and {Atayde}, L. and {Atek}, H. and {Atrio-Barandela}, F. and {Aubert}, M. and {Aubourg}, E. and {Auphan}, T. and {Auricchio}, N. and {Aussel}, B. and {Aussel}, H. and {Avelino}, P.~P. and {Avgoustidis}, A. and {Avila}, S. and {Awan}, S. and {Azzollini}, R. and {Baccigalupi}, C. and {Bachelet}, E. and {Bacon}, D. and {Baes}, M. and {Bagley}, M.~B. and {Bahr-Kalus}, B. and {Balaguera-Antolinez}, A. and {Balbinot}, E. and {Balcells}, M. and {Baldi}, M. and {Baldry}, I. and {Balestra}, A. and {Ballardini}, M. and {Ballester}, O. and {Balogh}, M. and {Ba{\~n}ados}, E. and {Barbier}, R. and {Bardelli}, S. and {Baron}, M. and {Barreiro}, T. and {Barrena}, R. and {Barriere}, J. -C. and {Barros}, B.~J. and {Barthelemy}, A. and {Bartolo}, N. and {Basset}, A. and {Battaglia}, P. and {Battisti}, A.~J. and {Baugh}, C.~M. and {Baumont}, L. and {Bazzanini}, L. and {Beaulieu}, J. -P. and {Beckmann}, V. and {Belikov}, A.~N. and {Bel}, J. and {Bellagamba}, F. and {Bella}, M. and {Bellini}, E. and {Benabed}, K. and {Bender}, R. and {Benevento}, G. and {Bennett}, C.~L. and {Benson}, K. and {Bergamini}, P. and {Bermejo-Climent}, J.~R. and {Bernardeau}, F. and {Bertacca}, D. and {Berthe}, M. and {Berthier}, J. and {Bethermin}, M. and {Beutler}, F. and {Bevillon}, C. and {Bhargava}, S. and {Bhatawdekar}, R. and {Bianchi}, D. and {Bisigello}, L. and {Biviano}, A. and {Blake}, R.~P. and {Blanchard}, A. and {Blazek}, J. and {Blot}, L. and {Bosco}, A. and {Bodendorf}, C. and {Boenke}, T. and {B{\"o}hringer}, H. and {Boldrini}, P. and {Bolzonella}, M. and {Bonchi}, A. and {Bonici}, M. and {Bonino}, D. and {Bonino}, L. and {Bonvin}, C. and {Bon}, W. and {Booth}, J.~T. and {Borgani}, S. and {Borlaff}, A.~S. and {Borsato}, E. and {Bose}, B. and {Botticella}, M.~T. and {Boucaud}, A. and {Bouche}, F. and {Boucher}, J.~S. and {Boutigny}, D. and {Bouvard}, T. and {Bouwens}, R. and {Bouy}, H. and {Bowler}, R.~A.~A. and {Bozza}, V. and {Bozzo}, E. and {Branchini}, E. and {Brando}, G. and {Brau-Nogue}, S. and {Brekke}, P. and {Bremer}, M.~N. and {Brescia}, M. and {Breton}, M. -A. and {Brinchmann}, J. and {Brinckmann}, T. and {Brockley-Blatt}, C. and {Brodwin}, M. and {Brouard}, L. and {Brown}, M.~L. and {Bruton}, S. and {Bucko}, J. and {Buddelmeijer}, H. and {Buenadicha}, G. and {Buitrago}, F. and {Burger}, P. and {Burigana}, C. and {Busillo}, V. and {Busonero}, D. and {Cabanac}, R. and {Cabayol-Garcia}, L. and {Cagliari}, M.~S. and {Caillat}, A. and {Caillat}, L. and {Calabrese}, M. and {Calabro}, A. and {Calderone}, G. and {Calura}, F. and {Camacho Quevedo}, B. and {Camera}, S. and {Campos}, L. and {Ca{\~n}as-Herrera}, G. and {Candini}, G.~P. and {Cantiello}, M. and {Capobianco}, V. and {Cappellaro}, E. and {Cappelluti}, N. and {Cappi}, A. and {Caputi}, K.~I. and {Cara}, C. and {Carbone}, C. and {Cardone}, V.~F. and {Carella}, E. and {Carlberg}, R.~G. and {Carle}, M. and {Carminati}, L. and {Caro}, F. and {Carrasco}, J.~M. and {Carretero}, J. and {Carrilho}, P. and {Carron Duque}, J. and {Carry}, B.},
        title = "{Euclid: I. Overview of the Euclid mission}",
      journal = {\aap},
         year = 2025,
        month = may,
       volume = {697},
          eid = {A1},
        pages = {A1},
          doi = {10.1051/0004-6361/202450810},
archivePrefix = {arXiv},
       eprint = {2405.13491},
 primaryClass = {astro-ph.CO},
       adsurl = {https://ui.adsabs.harvard.edu/abs/2025A&A...697A...1E}
}

@ARTICLE{Everett+:2022ApJS..258...15E,
       author = {{Everett}, S. and {Yanny}, B. and {Kuropatkin}, N. and {Huff}, E.~M. and {Zhang}, Y. and {Myles}, J. and {Masegian}, A. and {Elvin-Poole}, J. and {Allam}, S. and {Bernstein}, G.~M. and {Sevilla-Noarbe}, I. and {Splettstoesser}, M. and {Sheldon}, E. and {Jarvis}, M. and {Amon}, A. and {Harrison}, I. and {Choi}, A. and {Hartley}, W.~G. and {Alarcon}, A. and {S{\'a}nchez}, C. and {Gruen}, D. and {Eckert}, K. and {Prat}, J. and {Tabbutt}, M. and {Busti}, V. and {Becker}, M.~R. and {MacCrann}, N. and {Diehl}, H.~T. and {Tucker}, D.~L. and {Bertin}, E. and {Jeltema}, T. and {Drlica-Wagner}, A. and {Gruendl}, R.~A. and {Bechtol}, K. and {Carnero Rosell}, A. and {Abbott}, T.~M.~C. and {Aguena}, M. and {Annis}, J. and {Bacon}, D. and {Bhargava}, S. and {Brooks}, D. and {Burke}, D.~L. and {Carrasco Kind}, M. and {Carretero}, J. and {Castander}, F.~J. and {Conselice}, C. and {Costanzi}, M. and {da Costa}, L.~N. and {Pereira}, M.~E.~S. and {De Vicente}, J. and {DeRose}, J. and {Desai}, S. and {Eifler}, T.~F. and {Evrard}, A.~E. and {Ferrero}, I. and {Fosalba}, P. and {Frieman}, J. and {Garc{\'\i}a-Bellido}, J. and {Gaztanaga}, E. and {Gerdes}, D.~W. and {Gutierrez}, G. and {Hinton}, S.~R. and {Hollowood}, D.~L. and {Honscheid}, K. and {Huterer}, D. and {James}, D.~J. and {Kent}, S. and {Krause}, E. and {Kuehn}, K. and {Lahav}, O. and {Lima}, M. and {Lin}, H. and {Maia}, M.~A.~G. and {Marshall}, J.~L. and {Melchior}, P. and {Menanteau}, F. and {Miquel}, R. and {Mohr}, J.~J. and {Morgan}, R. and {Muir}, J. and {Ogando}, R.~L.~C. and {Palmese}, A. and {Paz-Chinch{\'o}n}, F. and {Plazas}, A.~A. and {Rodriguez-Monroy}, M. and {Romer}, A.~K. and {Roodman}, A. and {Sanchez}, E. and {Scarpine}, V. and {Serrano}, S. and {Smith}, M. and {Soares-Santos}, M. and {Suchyta}, E. and {Swanson}, M.~E.~C. and {Tarle}, G. and {To}, C. and {Troxel}, M.~A. and {Varga}, T.~N. and {Weller}, J. and {Wilkinson}, R.~D. and {Wilkinson}, R.~D.},
        title = "{Dark Energy Survey Year 3 Results: Measuring the Survey Transfer Function with Balrog}",
      journal = {\apjs},
         year = 2022,
        month = jan,
       volume = {258},
       number = {1},
          eid = {15},
        pages = {15},
          doi = {10.3847/1538-4365/ac26c1},
archivePrefix = {arXiv},
       eprint = {2012.12825},
 primaryClass = {astro-ph.CO},
       adsurl = {https://ui.adsabs.harvard.edu/abs/2022ApJS..258...15E}
}

@ARTICLE{Chaussidon+:2025JCAP...06..029C,
       author = {{Chaussidon}, E. and {Y{\`e}che}, C. and {de Mattia}, A. and {Payerne}, C. and {McDonald}, P. and {Ross}, A.~J. and {Ahlen}, S. and {Bianchi}, D. and {Brooks}, D. and {Burtin}, E. and {Claybaugh}, T. and {de la Macorra}, A. and {Doel}, P. and {Ferraro}, S. and {Font-Ribera}, A. and {Forero-Romero}, J.~E. and {Gazta{\~n}aga}, E. and {Gil-Mar{\'\i}n}, H. and {Gontcho}, S. Gontcho A. and {Gutierrez}, G. and {Guy}, J. and {Honscheid}, K. and {Howlett}, C. and {Huterer}, D. and {Kehoe}, R. and {Kirkby}, D. and {Kisner}, T. and {Kremin}, A. and {Le Guillou}, L. and {Levi}, M.~E. and {Manera}, M. and {Meisner}, A. and {Miquel}, R. and {Moustakas}, J. and {Newman}, J.~A. and {Niz}, G. and {Palanque-Delabrouille}, N. and {Percival}, W.~J. and {Prada}, F. and {P{\'e}rez-R{\`a}fols}, I. and {Ravoux}, C. and {Rossi}, G. and {Sanchez}, E. and {Schlegel}, D. and {Schubnell}, M. and {Seo}, H. and {Sprayberry}, D. and {Tarl{\'e}}, G. and {Vargas-Maga{\~n}a}, M. and {Weaver}, B.~A. and {Zhao}, C. and {Zou}, H.},
        title = "{Constraining primordial non-Gaussianity with DESI 2024 LRG and QSO samples}",
      journal = {\jcap},
         year = 2025,
        month = jun,
       volume = {2025},
       number = {6},
          eid = {029},
        pages = {029},
          doi = {10.1088/1475-7516/2025/06/029},
archivePrefix = {arXiv},
       eprint = {2411.17623},
 primaryClass = {astro-ph.CO},
       adsurl = {https://ui.adsabs.harvard.edu/abs/2025JCAP...06..029C}
}

@ARTICLE{Cagliari+:2024JCAP...08..036C,
       author = {{Cagliari}, Marina S. and {Castorina}, Emanuele and {Bonici}, Marco and {Bianchi}, Davide},
        title = "{Optimal constraints on Primordial non-Gaussianity with the eBOSS DR16 quasars in Fourier space}",
      journal = {\jcap},
         year = 2024,
        month = aug,
       volume = {2024},
       number = {8},
          eid = {036},
        pages = {036},
          doi = {10.1088/1475-7516/2024/08/036},
archivePrefix = {arXiv},
       eprint = {2309.15814},
 primaryClass = {astro-ph.CO},
       adsurl = {https://ui.adsabs.harvard.edu/abs/2024JCAP...08..036C}
}

@ARTICLE{PlanckCollaboration:2020A&A...641A...9P,
       author = {{Planck Collaboration} and {Akrami}, Y. and {Arroja}, F. and {Ashdown}, M. and {Aumont}, J. and {Baccigalupi}, C. and {Ballardini}, M. and {Banday}, A.~J. and {Barreiro}, R.~B. and {Bartolo}, N. and {Basak}, S. and {Benabed}, K. and {Bernard}, J. -P. and {Bersanelli}, M. and {Bielewicz}, P. and {Bond}, J.~R. and {Borrill}, J. and {Bouchet}, F.~R. and {Bucher}, M. and {Burigana}, C. and {Butler}, R.~C. and {Calabrese}, E. and {Cardoso}, J. -F. and {Casaponsa}, B. and {Challinor}, A. and {Chiang}, H.~C. and {Colombo}, L.~P.~L. and {Combet}, C. and {Crill}, B.~P. and {Cuttaia}, F. and {de Bernardis}, P. and {de Rosa}, A. and {de Zotti}, G. and {Delabrouille}, J. and {Delouis}, J. -M. and {Di Valentino}, E. and {Diego}, J.~M. and {Dor{\'e}}, O. and {Douspis}, M. and {Ducout}, A. and {Dupac}, X. and {Dusini}, S. and {Efstathiou}, G. and {Elsner}, F. and {En{\ss}lin}, T.~A. and {Eriksen}, H.~K. and {Fantaye}, Y. and {Fergusson}, J. and {Fernandez-Cobos}, R. and {Finelli}, F. and {Frailis}, M. and {Fraisse}, A.~A. and {Franceschi}, E. and {Frolov}, A. and {Galeotta}, S. and {Galli}, S. and {Ganga}, K. and {G{\'e}nova-Santos}, R.~T. and {Gerbino}, M. and {Gonz{\'a}lez-Nuevo}, J. and {G{\'o}rski}, K.~M. and {Gratton}, S. and {Gruppuso}, A. and {Gudmundsson}, J.~E. and {Hamann}, J. and {Handley}, W. and {Hansen}, F.~K. and {Herranz}, D. and {Hivon}, E. and {Huang}, Z. and {Jaffe}, A.~H. and {Jones}, W.~C. and {Jung}, G. and {Keih{\"a}nen}, E. and {Keskitalo}, R. and {Kiiveri}, K. and {Kim}, J. and {Krachmalnicoff}, N. and {Kunz}, M. and {Kurki-Suonio}, H. and {Lamarre}, J. -M. and {Lasenby}, A. and {Lattanzi}, M. and {Lawrence}, C.~R. and {Le Jeune}, M. and {Levrier}, F. and {Lewis}, A. and {Liguori}, M. and {Lilje}, P.~B. and {Lindholm}, V. and {L{\'o}pez-Caniego}, M. and {Ma}, Y. -Z. and {Mac{\'\i}as-P{\'e}rez}, J.~F. and {Maggio}, G. and {Maino}, D. and {Mandolesi}, N. and {Marcos-Caballero}, A. and {Maris}, M. and {Martin}, P.~G. and {Mart{\'\i}nez-Gonz{\'a}lez}, E. and {Matarrese}, S. and {Mauri}, N. and {McEwen}, J.~D. and {Meerburg}, P.~D. and {Meinhold}, P.~R. and {Melchiorri}, A. and {Mennella}, A. and {Migliaccio}, M. and {Miville-Desch{\^e}nes}, M. -A. and {Molinari}, D. and {Moneti}, A. and {Montier}, L. and {Morgante}, G. and {Moss}, A. and {M{\"u}nchmeyer}, M. and {Natoli}, P. and {Oppizzi}, F. and {Pagano}, L. and {Paoletti}, D. and {Partridge}, B. and {Patanchon}, G. and {Perrotta}, F. and {Pettorino}, V. and {Piacentini}, F. and {Polenta}, G. and {Puget}, J. -L. and {Rachen}, J.~P. and {Racine}, B. and {Reinecke}, M. and {Remazeilles}, M. and {Renzi}, A. and {Rocha}, G. and {Rubi{\~n}o-Mart{\'\i}n}, J.~A. and {Ruiz-Granados}, B. and {Salvati}, L. and {Savelainen}, M. and {Scott}, D. and {Shellard}, E.~P.~S. and {Shiraishi}, M. and {Sirignano}, C. and {Sirri}, G. and {Smith}, K. and {Spencer}, L.~D. and {Stanco}, L. and {Sunyaev}, R. and {Suur-Uski}, A. -S. and {Tauber}, J.~A. and {Tavagnacco}, D. and {Tenti}, M. and {Toffolatti}, L. and {Tomasi}, M. and {Trombetti}, T. and {Valiviita}, J. and {Van Tent}, B. and {Vielva}, P. and {Villa}, F. and {Vittorio}, N. and {Wandelt}, B.~D. and {Wehus}, I.~K. and {Zacchei}, A. and {Zonca}, A.},
        title = "{Planck 2018 results. IX. Constraints on primordial non-Gaussianity}",
      journal = {\aap},
         year = 2020,
        month = sep,
       volume = {641},
          eid = {A9},
        pages = {A9},
          doi = {10.1051/0004-6361/201935891},
archivePrefix = {arXiv},
       eprint = {1905.05697},
 primaryClass = {astro-ph.CO},
       adsurl = {https://ui.adsabs.harvard.edu/abs/2020A&A...641A...9P}
}

@ARTICLE{Adame+:2025JCAP...07..017A,
       author = {{Adame}, A.~G. and {Aguilar}, J. and {Ahlen}, S. and {Alam}, S. and {Alexander}, D.~M. and {Alvarez}, M. and {Alves}, O. and {Anand}, A. and {Andrade}, U. and {Armengaud}, E. and {Avila}, S. and {Aviles}, A. and {Awan}, H. and {Bailey}, S. and {Baltay}, C. and {Bault}, A. and {Behera}, J. and {BenZvi}, S. and {Beutler}, F. and {Bianchi}, D. and {Blake}, C. and {Blum}, R. and {Brieden}, S. and {Brodzeller}, A. and {Brooks}, D. and {Brown}, Z. and {Buckley-Geer}, E. and {Burtin}, E. and {Calderon}, R. and {Canning}, R. and {Carnero Rosell}, A. and {Cereskaite}, R. and {Cervantes-Cota}, J.~L. and {Chabanier}, S. and {Chaussidon}, E. and {Chaves-Montero}, J. and {Chen}, S. and {Chen}, X. and {Claybaugh}, T. and {Cole}, S. and {Cuceu}, A. and {Davis}, T.~M. and {Dawson}, K. and {de la Macorra}, A. and {de Mattia}, A. and {Deiosso}, N. and {Demina}, R. and {Dey}, A. and {Dey}, B. and {Ding}, Z. and {Doel}, P. and {Edelstein}, J. and {Eftekharzadeh}, S. and {Eisenstein}, D.~J. and {Elliott}, A. and {Fagrelius}, P. and {Fanning}, K. and {Ferraro}, S. and {Ereza}, J. and {Findlay}, N. and {Flaugher}, B. and {Font-Ribera}, A. and {Forero-S{\'a}nchez}, D. and {Forero-Romero}, J.~E. and {Frenk}, C.~S. and {Garcia-Quintero}, C. and {Gazta{\~n}aga}, E. and {Gil-Mar{\'\i}n}, H. and {Gontcho}, S. Gontcho A. and {Gonzalez-Morales}, A.~X. and {Gonzalez-Perez}, V. and {Gordon}, C. and {Green}, D. and {Gruen}, D. and {Gsponer}, R. and {Gutierrez}, G. and {Guy}, J. and {Hadzhiyska}, B. and {Hahn}, C. and {Hanif}, M.~M.~S. and {Herrera-Alcantar}, H.~K. and {Honscheid}, K. and {Hou}, J. and {Howlett}, C. and {Huterer}, D. and {Ir{\v{s}}i{\v{c}}}, V. and {Ishak}, M. and {Juneau}, S. and {Kara{\c{c}}ayl{\i}}, N.~G. and {Kehoe}, R. and {Kent}, S. and {Kirkby}, D. and {Kitaura}, F. -S. and {Kong}, H. and {Kremin}, A. and {Krolewski}, A. and {Lai}, Y. and {Lan}, T. -W. and {Landriau}, M. and {Lang}, D. and {Lasker}, J. and {Le Goff}, J.~M. and {Le Guillou}, L. and {Leauthaud}, A. and {Levi}, M.~E. and {Li}, T.~S. and {Lodha}, K. and {Magneville}, C. and {Manera}, M. and {Margala}, D. and {Martini}, P. and {Maus}, M. and {McDonald}, P. and {Medina-Varela}, L. and {Meisner}, A. and {Mena-Fern{\'a}ndez}, J. and {Miquel}, R. and {Moon}, J. and {Moore}, S. and {Moustakas}, J. and {Mudur}, N. and {Mueller}, E. and {Mu{\~n}oz-Guti{\'e}rrez}, A. and {Myers}, A.~D. and {Nadathur}, S. and {Napolitano}, L. and {Neveux}, R. and {Newman}, J.~A. and {Nguyen}, N.~M. and {Nie}, J. and {Niz}, G. and {Noriega}, H.~E. and {Padmanabhan}, N. and {Paillas}, E. and {Palanque-Delabrouille}, N. and {Pan}, J. and {Penmetsa}, S. and {Percival}, W.~J. and {Pieri}, M.~M. and {Pinon}, M. and {Poppett}, C. and {Porredon}, A. and {Prada}, F. and {P{\'e}rez-Fern{\'a}ndez}, A. and {P{\'e}rez-R{\`a}fols}, I. and {Rabinowitz}, D. and {Raichoor}, A. and {Ram{\'\i}rez-P{\'e}rez}, C. and {Ramirez-Solano}, S. and {Rashkovetskyi}, M. and {Ravoux}, C. and {Rezaie}, M. and {Rich}, J. and {Rocher}, A. and {Rockosi}, C. and {Roe}, N.~A. and {Rosado-Marin}, A. and {Ross}, A.~J. and {Rossi}, G. and {Ruggeri}, R. and {Ruhlmann-Kleider}, V. and {Samushia}, L. and {Sanchez}, E. and {Saulder}, C. and {Schlafly}, E.~F. and {Schlegel}, D. and {Scholte}, D. and {Schubnell}, M. and {Seo}, H. and {Sharples}, R. and {Silber}, J. and {Slosar}, A. and {Smith}, A. and {Sprayberry}, D. and {Tan}, T. and {Tarl{\'e}}, G. and {Trusov}, S. and {Vaisakh}, R. and {Valcin}, D. and {Valdes}, F. and {Vargas-Maga{\~n}a}, M. and {Verde}, L. and {Walther}, M. and {Wang}, B. and {Wang}, M.~S. and {Weaver}, B.~A. and {Weaverdyck}, N. and {Wechsler}, R.~H. and {Weinberg}, D.~H. and {White}, M. and {Wilson}, M.~J. and {Yu}, J. and {Yu}, Y. and {Yuan}, S. and {Y{\`e}che}, C. and {Zaborowski}, E.~A. and {Zarrouk}, P. and {Zhang}, H. and {Zhao}, C. and {Zhao}, R.},
        title = "{DESI 2024 II: sample definitions, characteristics, and two-point clustering statistics}",
      journal = {\jcap},
         year = 2025,
        month = jul,
       volume = {2025},
       number = {7},
          eid = {017},
        pages = {017},
          doi = {10.1088/1475-7516/2025/07/017},
archivePrefix = {arXiv},
       eprint = {2411.12020},
 primaryClass = {astro-ph.CO},
       adsurl = {https://ui.adsabs.harvard.edu/abs/2025JCAP...07..017A}
}

@ARTICLE{Cagliari+:2025JCAP...07..043C,
       author = {{Cagliari}, Marina S. and {Barberi-Squarotti}, Matilde and {Pardede}, Kevin and {Castorina}, Emanuele and {D'Amico}, Guido},
        title = "{Bispectrum constraints on Primordial Non-Gaussianities with the eBOSS DR16 quasars}",
      journal = {\jcap},
         year = 2025,
        month = jul,
       volume = {2025},
       number = {7},
          eid = {043},
        pages = {043},
          doi = {10.1088/1475-7516/2025/07/043},
archivePrefix = {arXiv},
       eprint = {2502.14758},
 primaryClass = {astro-ph.CO},
       adsurl = {https://ui.adsabs.harvard.edu/abs/2025JCAP...07..043C}
}

@ARTICLE{Cabass+:2022PhRvD.106d3506C,
       author = {{Cabass}, Giovanni and {Ivanov}, Mikhail M. and {Philcox}, Oliver H.~E. and {Simonovi{\'c}}, Marko and {Zaldarriaga}, Matias},
        title = "{Constraints on multifield inflation from the BOSS galaxy survey}",
      journal = {\prd},
         year = 2022,
        month = aug,
       volume = {106},
       number = {4},
          eid = {043506},
        pages = {043506},
          doi = {10.1103/PhysRevD.106.043506},
archivePrefix = {arXiv},
       eprint = {2204.01781},
 primaryClass = {astro-ph.CO},
       adsurl = {https://ui.adsabs.harvard.edu/abs/2022PhRvD.106d3506C}
}

@ARTICLE{Reid+:2016MNRAS.455.1553R,
       author = {{Reid}, Beth and {Ho}, Shirley and {Padmanabhan}, Nikhil and {Percival}, Will J. and {Tinker}, Jeremy and {Tojeiro}, Rita and {White}, Martin and {Eisenstein}, Daniel J. and {Maraston}, Claudia and {Ross}, Ashley J. and {S{\'a}nchez}, Ariel G. and {Schlegel}, David and {Sheldon}, Erin and {Strauss}, Michael A. and {Thomas}, Daniel and {Wake}, David and {Beutler}, Florian and {Bizyaev}, Dmitry and {Bolton}, Adam S. and {Brownstein}, Joel R. and {Chuang}, Chia-Hsun and {Dawson}, Kyle and {Harding}, Paul and {Kitaura}, Francisco-Shu and {Leauthaud}, Alexie and {Masters}, Karen and {McBride}, Cameron K. and {More}, Surhud and {Olmstead}, Matthew D. and {Oravetz}, Daniel and {Nuza}, Sebasti{\'a}n E. and {Pan}, Kaike and {Parejko}, John and {Pforr}, Janine and {Prada}, Francisco and {Rodr{\'\i}guez-Torres}, Sergio and {Salazar-Albornoz}, Salvador and {Samushia}, Lado and {Schneider}, Donald P. and {Sc{\'o}ccola}, Claudia G. and {Simmons}, Audrey and {Vargas-Magana}, Mariana},
        title = "{SDSS-III Baryon Oscillation Spectroscopic Survey Data Release 12: galaxy target selection and large-scale structure catalogues}",
      journal = {\mnras},
         year = 2016,
        month = jan,
       volume = {455},
       number = {2},
        pages = {1553-1573},
          doi = {10.1093/mnras/stv2382},
archivePrefix = {arXiv},
       eprint = {1509.06529},
 primaryClass = {astro-ph.CO},
       adsurl = {https://ui.adsabs.harvard.edu/abs/2016MNRAS.455.1553R}
}

@ARTICLE{Foglieni+2023,
       author = {{Foglieni}, Matteo and {Pantiri}, Mattia and {Di Dio}, Enea and {Castorina}, Emanuele},
        title = "{Large Scale Limit of the Observed Galaxy Power Spectrum}",
      journal = {\prl},
         year = 2023,
        month = sep,
       volume = {131},
       number = {11},
          eid = {111201},
        pages = {111201},
          doi = {10.1103/PhysRevLett.131.111201},
archivePrefix = {arXiv},
       eprint = {2303.03142},
 primaryClass = {astro-ph.CO},
       adsurl = {https://ui.adsabs.harvard.edu/abs/2023PhRvL.131k1201F}
}

\clearpage
\appendix
\section{Testing Against Model Misspecification}
\label{app:A}
In this Appendix, we review our attempts to isolate and explain the systematic which seems to be present in the QSO sample's highest $\ell$ bin, $65 < \ell < 96$. Among the tests we considered were:
\begin{enumerate}
    \item cutting the off-diagonals from the data vector,
    \item changing the assumed $p$ value, and
    \item allowing a more flexible FoG model and fitting its parameters.
\end{enumerate}

\subsection{Off-Diagonals}
Unaccounted for radial inhomogeneities in the data would manifest as excess power in the off-diagonals of the SFB power spectrum. Such radial inhomogeneities, on the largest scales, could arise from , e.g., systematics or FoG model misspecification. As such, it is reasonable to try cutting the off-diagonals, which may be where problematic systematics disproportionately contribute power, and see how the posteriors change.

\begin{figure}
  \centering
  \incgraph{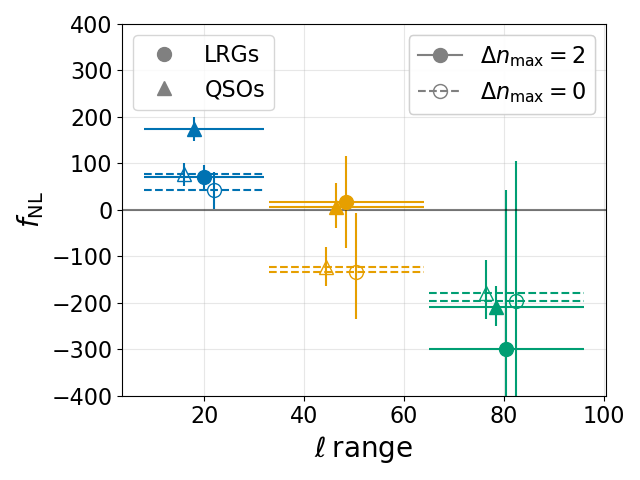}
  \caption{Inferred values of $\fnl$ with (filled points) and without (open points) the off-diagonals for the same $\ell$ bins considered in Figure \ref{fig:fnl_vs_binned_ell}. Colors are reused from Figure \ref{fig:fnl_vs_binned_ell}. When cutting the off-diagonals in the first $\ell$ bin, we gain the same benefits that we saw when cutting the $n=0$ mode: the QSO population comes into agreement with the LRG sample. This indicates that the systematic, which we hypothesize is stellar contamination, is isolated to not just the $n=0$ mode, but specifically the off-diagonal of the $n=0$ mode. Cutting the off-diagonals does not alleviate the QSOs largest $\ell$ bin's contention with zero and creates a tension in the middle $\ell$ bin with zero, perhaps an indication of offsetting systematics in the middle $\ell$ bin in the on- and off-diagonals.}
  \label{fig:vary_dn}
\end{figure}

Figure \ref{fig:vary_dn} shows the derived posteriors on $\fnl$ for the LRG and QSO samples with and without the off diagonals included.

We see that, when removing the off-diagonals ($\Delta n = 0$), the quasar's lowest $\ell$ bin comes into agreement with the LRG data points, indicating that the systematic in this bin, which we hypothesized in Section \ref{sec:binned_angular_scales} was stellar contamination, is specifically contaminating the off-diagonal modes of the SFB power spectrum.

Further, the middle $\ell$ bin, when removing the off-diagonals ($\Delta n = 0$), shifts downward for both LRGs (though with an error bar consistent with zero) and QSOs by approximately the same amount. This may be an indication of systematics in the on- and off-diagonals that are canceling each other out, akin to the discussion in Section \ref{sec:binned_angular_scales}.

Finally, the highest $\ell$ bin data points do not shift in a statistically significant way when removing the off-diagonals, indicating that the systematic does not reside in these modes.

\subsection{Changing the value of $p$}\label{sec:change-p}
The value of $p$, which, as described in Section \ref{sec:power_spectrum_theory}, characterizes the merger history of the tracer, controls the response of a tracer to $\fnl$ through $b_\phi$, which is in turn fully degenerate with the value of $\fnl$. QSOs, in previous studies, are typically prescribed a value of $p=1.6$, a divergence from the universality relation, reflecting that quasars, as a population average, have a recent merger. We test here the response of our inferred $\fnl$ to the choice in the value of $p$.

\begin{figure}
  \centering
  \incgraph{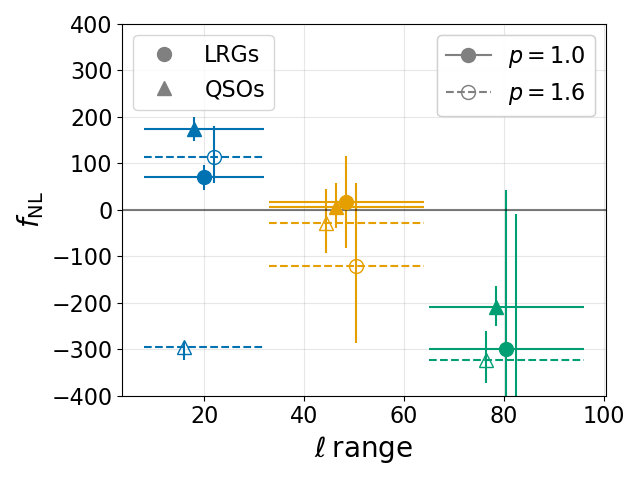}
  \caption{Inferred values of $\fnl$ when changing the value of $p$, which, in part, parametrizes the response of a sample to $\fnl$. Colors are reused from Figure \ref{fig:fnl_vs_binned_ell}. Filled points are used for $p=1$ and open points are used for $p=1.6$. The LRG sample is unaffected to within error bars in all $\ell$ bins. The QSOs lowest $\ell$ bin's inferred $\fnl$ depends strongly on the chosen value of $p$, but, as discussed in the text and further explored in Figure \ref{fig:fit_vs_fixed_bias}, we believe this is due to an ill-behaved bias fit, likely resulting from stellar contamination. The QSOs largest $\ell$ bin does not move toward zero significantly. The chosen value of $p$ certainly matters, but the effect is not large enough to explain the largest $\ell$ bin's discrepancy with $\fnl \sim 0$.}
  \label{fig:vary_p}
\end{figure}

We begin by simply changing the value of $p$ to 1.6 and redo the MCMC fitting for $\fnl$ and the bias parameters for both the LRGs and QSOs. The LRG $p$ value is not expected to be 1.6, but we vary it to test the dependence in the inferred value of $\fnl$ on $p$. We show in Figure \ref{fig:vary_p} the inferred $\fnl$ values for both populations. Focusing first on the LRG population, we do not see variation in the inferred value of $\fnl$ beyond the error bars when changing the value of $p$. Not so for the QSOs. In the lowest $\ell$ range bin, we see that our inferred value of $\fnl$ is shifted by about 500, from 200 to -300 and both points have very tight error bars. The middle $\ell$ range is consistent and the largest $\ell$ range has a very slight tension, but nothing close to the lowest $\ell$ bin.

We determined that this behavior is a complex interplay between the value chosen for $p$, the fit bias parameters, and the systematic that is afflicting this bin. In the data, the bias relation for the $8 < \ell < 32$ bin for both $p=1.0$ and $p=1.6$ evolves in the opposite direction of the rest of the bins, going from high bias values to low bias values as $z$ increases. We find that this is not the case for the EZMocks; the low $\ell$ bin sample for the EZMocks has a typical bias evolution consistent with that found in \citet{Cagliari+:2024JCAP...08..036C}.

\begin{figure}
  \centering
  \incgraph{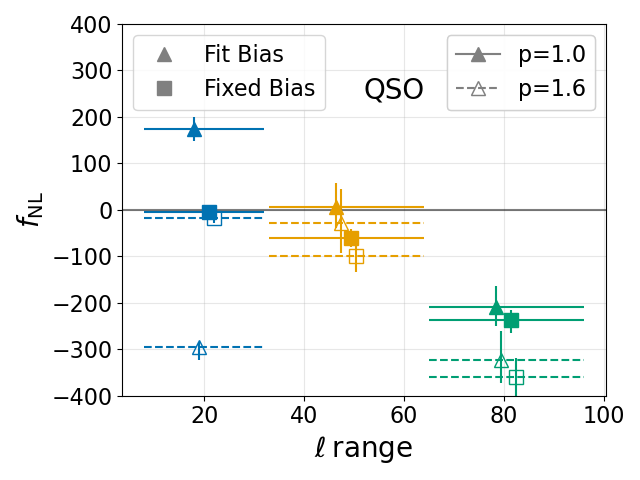}
  \caption{Similar to Figure \ref{fig:vary_p} but varying fixing versus fitting the bias parameters and only showing QSOs for clarity. Filled points indicate runs where we fit the bias and open points indicate runs where we fix the bias. When bias parameters are assigned instead of fit, the lowest $\ell$ bin is consistent with zero, highlighting that the ill-behaved bias parameters are responsible for the sporadic $\fnl$ behavior as a function of $p$ in Figure \ref{fig:vary_p}. Note that the bias must be fit in conjunction with $\fnl$ in a proper analysis.}
  \label{fig:fit_vs_fixed_bias}
\end{figure}

In Figure \ref{fig:fit_vs_fixed_bias}, we show the fitting results when we vary the value of $p$ and fit for the QSO bias versus fixing it at fiducial values to recreate the results from \citet{Cagliari+:2024JCAP...08..036C}. The lowest $\ell$ bin suddenly comes into agreement with $\fnl \sim 0$ when the bias is set instead of fit, indicating that the bias is not being fit well in this sample. Our hypothesis is that the stellar contamination afflicting this bin is rendering the fit bias values unphysical and that the inferred value of $\fnl$ can then evolve very quickly, depending on the choice of $p$, because the bias values cross the value of 1.6. As one can see in Eqs. \ref{eq:bias} and  \ref{eq:universal-bias}, when $b_1$ crosses the value of $p$, the physical response of clustering of the tracers to $\fnl$ changes sign.

While this makes interpretation of the lowest $\ell$ bin more challenging, it does not help explain the systematic in the $65 < \ell < 96$ bin. In fact, it made the discrepancy with $\fnl \sim 0$ more severe.

\subsection{FoG Modeling}\label{sec:fog-test}
Finally, we consider that the FoG model has insufficient complexity to describe the real behavior, which we hypothesize could lead to a bias in the inferred $\fnl$.

To test this, we allow more flexibility in the FoG model, letting it evolve in redshift in the same manner that our bias model evolves (with Chebyshev polynomials, see \cref{eq:sig_fogz_expansion}). With this new FoG model, we redo the inference, fitting for the bias parameters, $\fnl$, and the FoG parameters, for two sets of cuts: using $8 <\ell$ and $65 < \ell < 96$ for both the data and one randomly chosen EZmock. In the scenario that the FoG model was previously insufficiently complex and forced $\fnl$ to take on a very negative value, the $\fnl$ values of the $65 < \ell < 96$ data sample may see its inferred $\fnl$ moved toward zero as the FoG model changes to capture features in the data vector that were mimicking a strong negative $\fnl$ signal. The EZMocks and $8 < \ell$ samples will serve as some controls to better understand the behavior.

\begin{figure*}
  \centering
  \incgraph{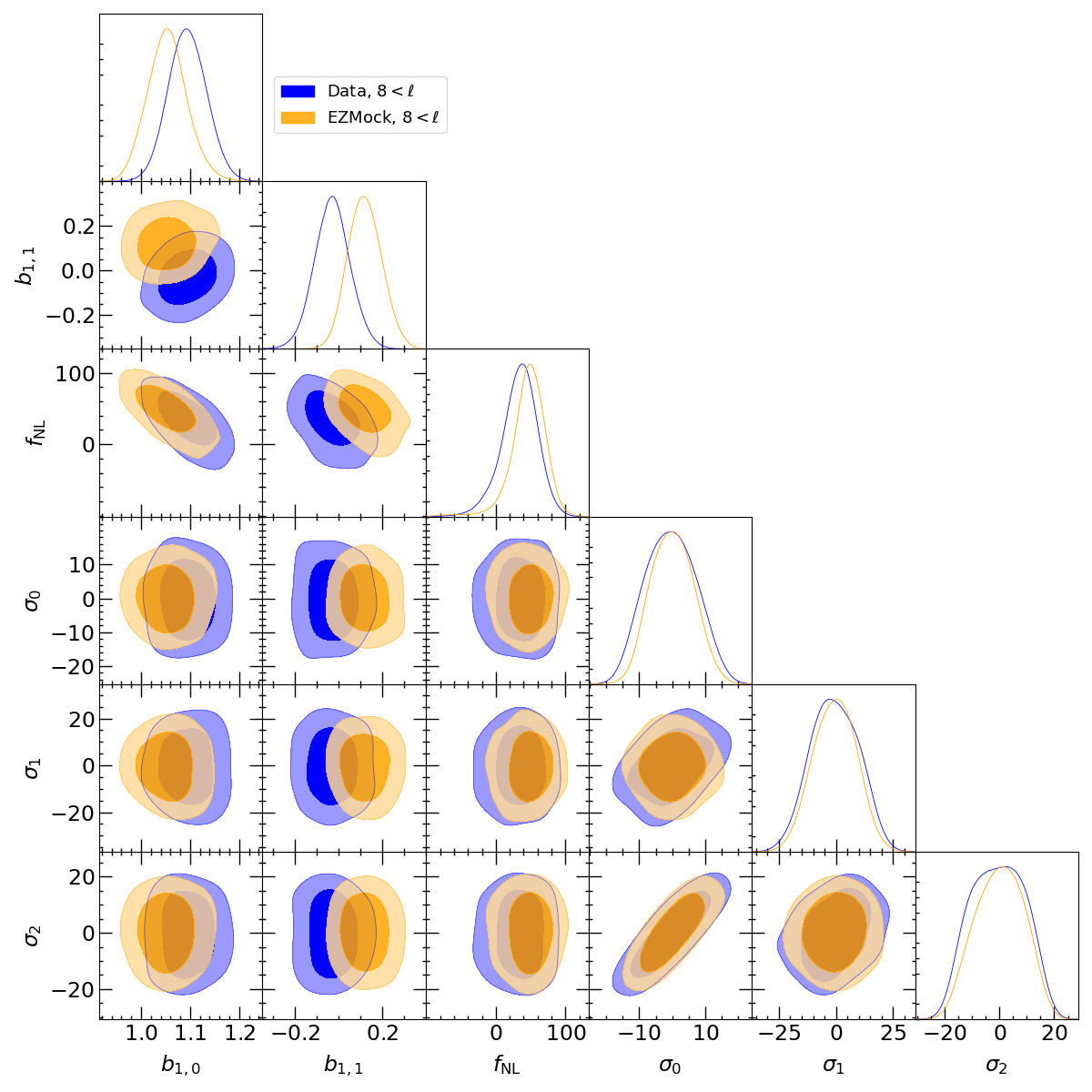}
  \caption{Corner plots for inferred biases, $\fnl$, and FoG parameters for the QSO sample (blue) and one randomly chosen EZmock (orange) both with $\ell_{min} = 8$. The most notable features are that 1) the FoG parameters agree well and 2) the bias parameters are in tension.}
  \label{fig:fog_full}
\end{figure*}

\begin{figure*}
  \centering
  \incgraph{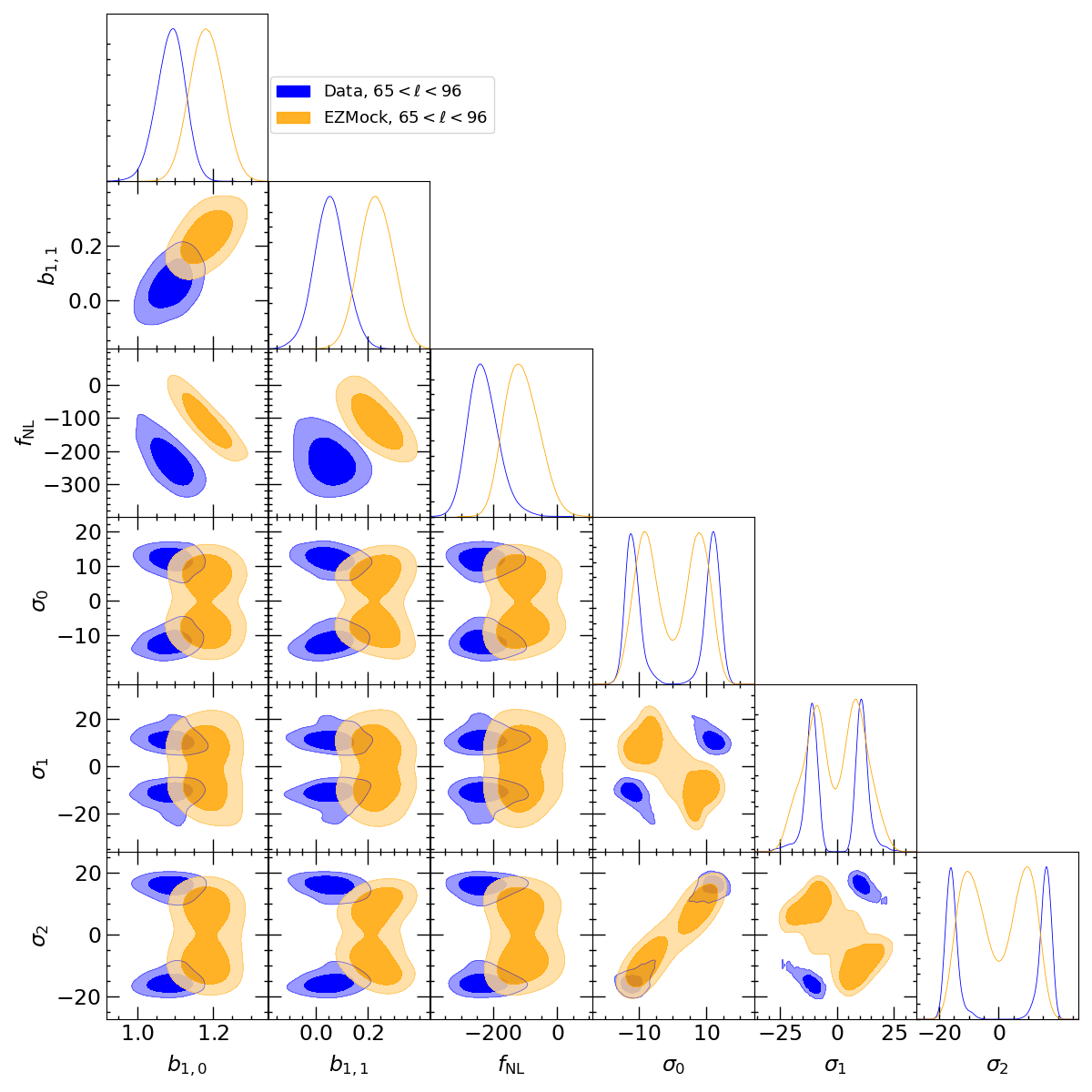}
  \caption{Similar to Figure \ref{fig:fog_full} except applying the cut $65 < \ell < 96$. Though the data favors FoG parameter values that are farther from zero than the EZMock, the $\fnl$ value still peaks at $\sim-250$, indicating that an insufficiently flexible FoG model was not attributable for the extremely negative $\fnl$ values in Figure \ref{fig:fnl_vs_binned_ell}.}
  \label{fig:fog_limited}
\end{figure*}

We show in Figures \ref{fig:fog_full} and \ref{fig:fog_limited} the corner plots for the $8 < \ell$ and $65 < \ell < 96$ samples, respectively, with the data in blue and the EZmock in orange. For the $8 < \ell$ sample (Figure \ref{fig:fog_full}), the $\fnl$ and FoG parameters all agree quite well. There is a difference in the fit bias parameters between the EZMock and the data, but this is not relevant to the test we are considering at the moment.

In the $65 < \ell < 96$ samples, Figure \ref{fig:fog_limited}, we see that the EZmock's $\fnl$ is fairly consistent with zero while the data still fits an $\fnl$ value of $\sim-250$. The FoG parameters between the EZMock and data are somewhat different; the data more strongly favors FoG parameter values away from zero. Note that the FoG parameters enter the model through a squared term (Section \ref{sec:sfb_formalism}), and so the positive and negative values for a given FoG parameter are equivalent, hence the symmetry across zero. Regardless, we do not see the value $\fnl$ change when we opened up the FoG fitting and allowed the model more evolution, so the extremely large negative $\fnl$ value is not being driven by inflexible FoG modeling.

\section{EZMock Inferred $\fnl$ Distribution Fits}
\label{app:B}
\begin{figure*}
  \centering
  \incgraph{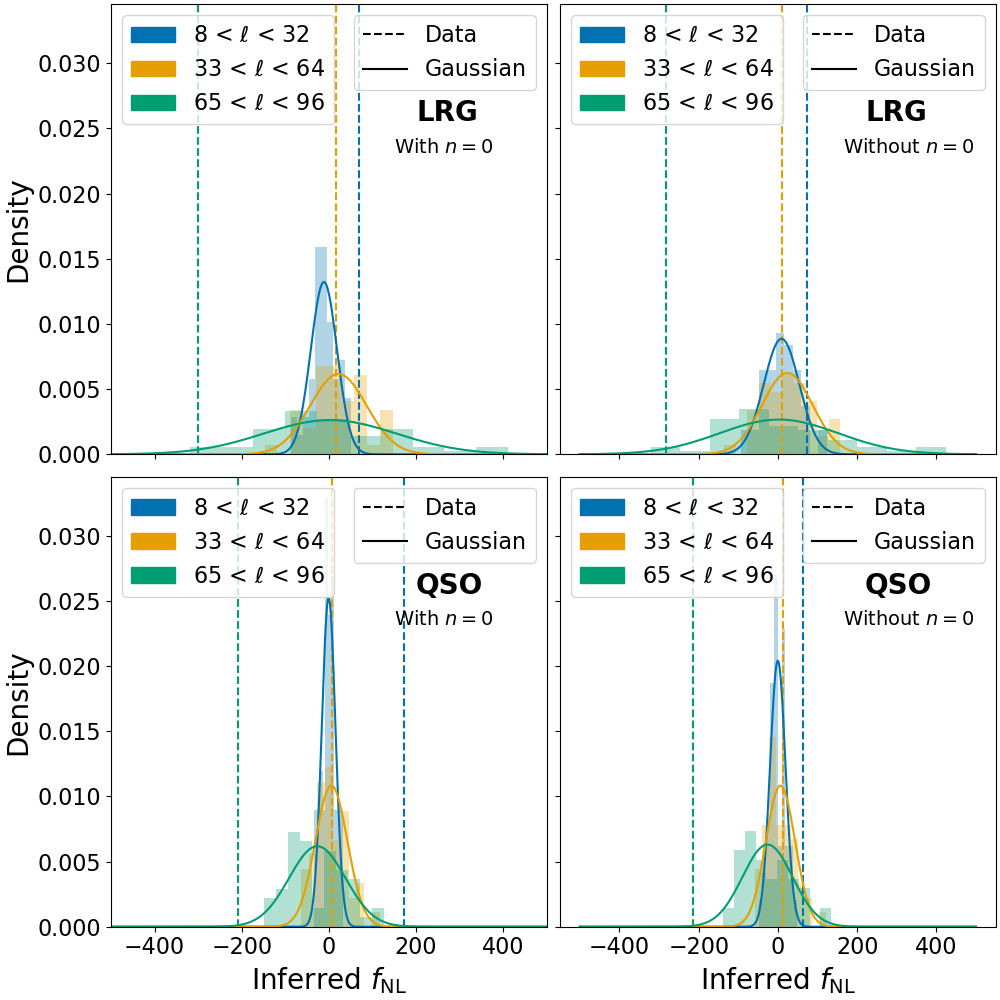}
  \caption{The distributions and Gaussian fits for the $N=50$ inferred $\fnl$ values from the EZMocks with the matching cuts to Figure \ref{fig:fnl_vs_binned_ell}. We test for inconsistency with normality in the EZMock distributions with the Shapiro-Wilk (SW) and Anderson-Darling (AD) tests. We find that none of the distributions show detectable deviations from normality (SW $\in [0.17, 0.94]$ and AD statistic $< AD_{crit,0.05}$ for all distributions).}
  \label{fig:app_ezmock_distributions}
\end{figure*}

In this Appendix, we present the Gaussian fits to the N=50 inferred $\fnl$ values from the EZMocks with the various cuts in Figure \ref{fig:fnl_vs_binned_ell}. Figure \ref{fig:app_ezmock_distributions} shows the distributions, none of which are found to be inconsistent with normality.

\ignore{Spherical Bessel functions and spherical harmonics satisfy orthogonality
relations
\ba
\label{eq:jljlDelta}
\delta^D(k-k')
&= \frac{2kk'}{\pi}\int_0^\infty\dd{r}\,r^2\,j_\ell(kr)\,j_\ell(k'r)\,, \\
\label{eq:YlmYlmDelta}
\delta^K_{\ell\ell'}\delta^K_{mm'}
&= \int\dd{\Omega}_{\rhat}\,Y_{\ell m}(\rhat)\,Y^*_{\ell'm'}(\rhat)\,.
\ea

The Laplacian in spherical coordinates is
\begin{align}
    \nabla^2f &= \frac{1}{r^2}\,\frac{\partial}{\partial r}\left(r^2\,\frac{\partial f}{\partial r}\right)
    + \frac{1}{r^2\sin\theta}\,\frac{\partial}{\partial\theta}\left(\sin\theta\,\frac{\partial f}{\partial\theta}\right)
    \nonumber \\
    &\quad
    + \frac{1}{r^2\sin^2\theta}\,\frac{\partial^2f}{\partial\phi^2}\,.
    \label{eq:laplacian_spherical}
\end{align}

The SFB transform pair is
\ba
\delta(\vr) &= \int\dd k\,\sum_{\ell m}
\left[\sqrt{\frac{2}{\pi}}\,k\,j_\ell(kr)\,Y_{\ell m}(\theta,\phi)\right]
\delta_{\ell m}(k)\,,
\label{eq:sfb_fourier_pair_a}
\\
\delta_{\ell m}(k) &=
\int\dd^3r
\left[\sqrt{\frac{2}{\pi}}\,k\,j_\ell(kr)\,Y^*_{\ell m}(\rhat)\right]
\delta(\vr)\,.
\label{eq:sfb_fourier_pair_b}
\ea}


\end{document}